\documentclass[preprint,aps,prd,superscriptaddress,nofootinbib]{revtex4-1}%

\usepackage{amsmath}
\usepackage{amssymb}
\usepackage{graphicx}% Include figure files
\usepackage{dcolumn}% Align table columns on decimal point
\usepackage{bm}% bold math
\usepackage{hyperref}% add hypertext capabilities
\usepackage{epsfig}
\usepackage{mathrsfs}
\usepackage{slashed}
\def\sl{\!\!\!/}
\def\g5{\gamma^5}

\def\e{\epsilon}
\def\ve{\varepsilon}

\def\beqn{\begin{eqnarray}}
\def\eeqn{\end{eqnarray}}
\def\beq{\begin{equation}}
\def\eeq{\end{equation}}

\def\bs{\boldsymbol}

\def\nn{\nonumber}

\begin{document}
\preprint{}

%%%%%%%%%%%%%%%%%%%%%%%%%%%%%%%%%%%%%%%%%%%%
\title{Analyticity, renormalization, and evolution\\ of the soft-quark 
function}
% Force line breaks with \\
%%%%%%%%%%%%%%%%%%%%%%%%%%%%%%%%%%%%%%%%%%%%
\author{Geoffrey~T.~Bodwin}
\email[]{gtb@anl.gov}
\affiliation{High Energy Physics Division, Argonne National Laboratory,
Argonne, Illinois 60439, USA}
\author{June-Haak~Ee}
\email[]{june\_haak\_ee@fudan.edu.cn}
\affiliation{Department of Physics, Korea University, Seoul 02841, Korea}
\affiliation{Key Laboratory of Nuclear Physics and Ion-beam Application (MOE) and Institute of Modern Physics, Fudan University, Shanghai 200433, China}
\author{Jungil~Lee}
\email[]{jungil@korea.ac.kr}
\affiliation{Department of Physics, Korea University, Seoul 02841, Korea}
\author{Xiang-Peng~Wang}
\email[]{xiangpeng.wang@anl.gov}
\affiliation{High Energy Physics Division, Argonne National Laboratory,
Argonne, Illinois 60439, USA}

%%%%%%%%%%%%%%%%%%%%%%%%%%%%%%%%%%%%%%%%%%%%
\date{\today}% It is always \today, today,
\begin{abstract}
We compute the renormalization and evolution of the soft-quark
    function that appears in the factorization theorem for Higgs-boson
    decays to two photons through a $b$-quark loop. Our computation
    confirms a conjecture by Liu, Mecaj, Neubert, Wang, and Fleming for
    the form of the renormalization and evolution of the soft-quark
    function in order $\alpha_s$. We also work out the analyticity
    structure of the soft-quark function by making use of light-cone
    perturbation theory.
\end{abstract}
\maketitle
%\tableofcontents
%%%%%%%%%%%%%%%%%%%%%%%%%%%%%%%%%%%%%%%%%%%%

\section{Introduction}

One of the principal decay modes of the Higgs boson is the decay to two
photons ($H\to \gamma\gamma$). Comparison of the theoretical prediction
for the rate of this decay mode with experimental measurements
provides an important test of the standard model, and improvements in
the precision of the theoretical prediction can lead to increasingly
stringent tests.

\begin{figure}
\begin{center}
\includegraphics[width=0.25\columnwidth]{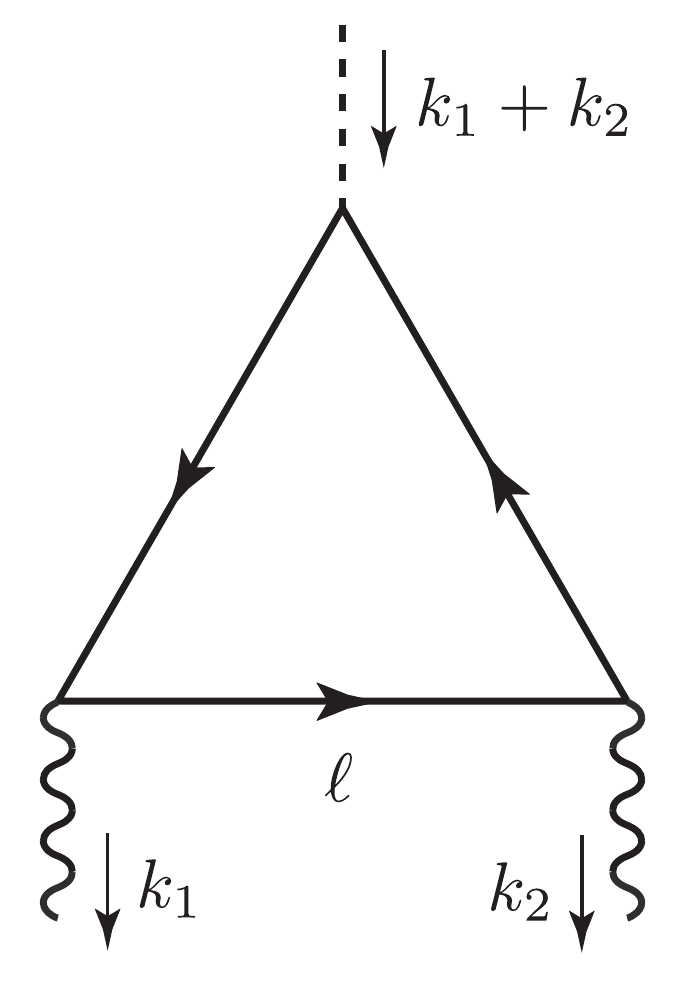}
\caption{$H\rightarrow b\bar{b}\rightarrow \gamma\gamma$ at leading order. 
The dashed line represents the Higgs boson, the solid line represents
    the $b$ quark, and the wavy lines are the photons.
\label{fig:H2gamma}%
}
\end{center}
\end{figure}

One mechanism for this decay mode proceeds through the coupling
of the Higgs boson to a virtual $b$-quark loop, which, in turn,
couples to the final-state photons (Fig.~\ref{fig:H2gamma}). While this
is not the dominant mechanism in $H\to\gamma\gamma$ decays, it is
relevant to precision calculations of the decay rate. Furthermore, as we
will explain, it is of particular theoretical interest.

Perturbative-QCD corrections to $H\to \gamma\gamma$ through a
    $b$-quark loop contain logarithms of $m_H^2/m_b^2$
    \cite{Spira:1995rr}, where $m_H$ and $m_b$ are the Higgs-boson and
    $b$-quark masses, respectively.  Resummation of these large
    logarithms is essential to a well-controlled theoretical
    prediction. A traditional approach to resummation would be to make
    use of the $b\bar b$ light-cone distribution amplitudes for the
    photon.  However, such an approach fails in this case because the
    amplitude for the decay process is proportional to $m_b$ at the
    leading nontrivial order in $m_b/m_H$. As is well known, such
    helicity-flip processes contain endpoint singularities that arise
    when all of the momentum of a spectator $b$ quark or antiquark is
    transferred to an active $b$ quark or antiquark. The endpoint
    singularities result in ill-defined quantities when one attempts
        to apply traditional resummation methods.

The endpoint singularities for exclusive amplitudes involving heavy
quarks have been known for some time
\cite{Beneke:2000ry,Beneke:2001at,Beneke:2001ev,Beneke:2003pa,
  Beneke:2003zv,Jia:2010fw,Benzke:2010js}.  They are associated with
     amplitudes that are suppressed by a power of the large momentum
           transfer and correspond to a pinch-singular region in
     momentum space in which the heavy quark carries a soft momentum
     \cite{Bodwin:2014dqa}.\footnote{Corrections at subleading power
           in the inverse of the large momentum transfer have also been
           discussed in the context of inclusive cross sections.  See,
           for example Refs.~\cite{Beneke:2019oqx,Moult:2019mog,Moult:2019uhz,vanBeekveld:2019prq}.}
         Making use of this insight, the authors of
         Ref.~\cite{Liu:2019oav} have proposed a factorization theorem
         for $H\to\gamma\gamma$ through a $b$-quark loop that decomposes
         the endpoint contributions into the convolution of a soft-quark
         function with jet functions that account for contributions that
         arise from collinear quarks and gluons.  In this factorization
         theorem, the endpoint contributions are well defined, and it
         can be used to resum logarithms of
         $m_H^2/m_b^2$.\footnote{Resummation of leading single and
         double logarithms in $H\to\gamma\gamma$ through a $b$-quark
         loop has also been considered in
         Refs.~\cite{Kotsky:1997rq,Akhoury:2001mz}. However, it is not
         clear that the methods in these papers can be generalized
         beyond the level of the leading single- and double-logarithm
         approximations.}  A similar factorization theorem has been
         proposed in Ref.~\cite{Wang:2019mym}.  The renormalized form of
         the factorization theorem has been given in
         Refs.~\cite{Liu:2020tzd,Liu:2020wbn}.  The factorization
         theorem is stated in the language of soft-collinear effective
         theory (SCET)
         \cite{Bauer:2000yr,Bauer:2001yt,Beneke:2002ni,Bauer:2002nz,Beneke:2002ph}.

As we have mentioned, one of the elements of the factorization theorem
is a soft-quark function. (Hereinafter, we refer to it as the ``soft
function.'')  Its renormalization-group evolution equation is an
      essential component in the resummation of logarithms of
      $m_{H}^2/m_b^2$, and can be derived, to a given order in
      $\alpha_s$, once one has worked out the renormalization condition
      for the soft function to that order in $\alpha_s$.

Note that, in order to derive the evolution equation, one must work
    out the renormalization condition for a generic soft function. In
          deriving the evolution equation, it 
          is not sufficient to compute the UV divergences of the
          fixed-order (in $\alpha_s$) soft function, because the
          renormalization condition for the soft function involves a
          convolution of the renormalization factor with the soft
          function, rather than a simple multiplication. One must make
          the convolution integrals explicit in order to deduce the
          renormalization condition.

In Ref.~\cite{Liu:2019oav}, the soft function was computed at
      order-$\alpha_s$, and the UV poles in dimensional regularization
      were identified. However, as we have mentioned, such a calculation
      is insufficient to work out the evolution equation for the soft
      function.

In Ref.~\cite{Liu:2020eqe}, a conjecture was given for the
    renormalization/evolution of the soft function through order
    $\alpha_s^2$. In that work, the renormalization condition for the
    soft function was derived by assuming the consistency condition,
          which implies the renormalization-group invariance of the
    so-called ``soft sector'' of the factorization theorem, which
    consists of the product of a certain Wilson coefficient with the
    convolution of the soft function with radiative jet functions
    \cite{Liu:2019oav}.  However, the renormalization-group invariance
    of the soft sector has not been established, and, indeed, in
    Ref.~\cite{Liu:2019oav}, it is pointed out that the
    renormalization-group invariance is violated at one-loop level once
    one has imposed rapidity regulators that are needed to make the
    convolution integral in the soft sector well defined. The
          conjecture for the evolution equation for the soft function is
          also stated, without further explanation, in
          Ref.~\cite{Liu:2020wbn}

Given the importance of the soft function in the
factorization/resummation program, it is essential to put the
renormalization condition for the soft function on a more solid
footing.  In the present paper, we work out the
    renormalization/evolution of the soft function through order
$\alpha_s$. Our computation confirms the conjecture in
Ref.~\cite{Liu:2020eqe} through order $\alpha_s$. The analysis is
    novel because, as we will explain in detail, the momentum routing in
    the soft function is unorthodox: The various components of the loop
    momenta route through different propagators and vertices. This
    unorthodox momentum arises as a consequence of the factorization of
    the soft function from the radiative jet functions. It results in
    some unexpected analyticity properties of the soft function. It
          also leads to a nonlocal renormalization condition for the
          soft function, although, as is well known, nonlocal
          renormalization conditions can also appear in the case of
          standard momentum routing, for example, in the renormalization
          of parton distributions.

Our calculation relies on the analytic structure of the soft function in
the complex plane of its longitudinal-momentum variables. In
Ref.~\cite{Liu:2019oav}, the analyticity properties of the soft function
were asserted. In the present paper, we establish those properties by
showing how the regions of non-analyticity arise in specific examples
and by giving a general argument for the analyticity everywhere else
in the complex plane.  Our arguments make use of light-front
perturbation theory.

The remainder of this paper is organized as follows. In
Sec.~\ref{sec:notation-defs}, we define some of our notation.  Section
\ref{sec:factorization-thm} contains a statement of the factorization
theorem for $H\to \gamma\gamma$ through a $b$-quark loop. In
Sec.~\ref{sec:soft-op-defn} we give the operator definition of the soft
function, and in Sec.~\ref{sec:struct-fns} we give its decomposition
into structure functions and define its discontinuity. We present the
diagrammatic form of the soft function in
Sec.~\ref{sec:diagrammatic-soft-fn}, discuss the leading-order (LO) and
next-to-leading order (NLO) contributions to the soft function in
Sec.~\ref{sec:LO-NLO-soft-fn}, and examine the analyticity of the
soft function in Sec.~\ref{sec:analyticity}. Examples that illustrate
the analyticity structure of the soft function and details of the
general analyticity argument are given in the Appendix.
Sec.~\ref{sec:renorm-soft-fn} contains a general discussion of the
renormalization of the soft function. In Sec.~\ref{sec:one-loop-calc},
we present our calculation of the one-loop renormalization of the soft
function and write down the evolution equation for the particular
    structure function that appears in the process $H\to
    \gamma\gamma$. Finally, we summarize our results in
Sec.~\ref{sec:summary}.

\section{Notation and conventions\label{sec:notation-defs}}

In this section, we establish some of our notation and conventions,
which generally agree with those that are used in
Ref.~\cite{Liu:2019oav}.

In Fig.~\ref{fig:H2gamma} we show the LO Feynman diagram for the decay
    $H\to\gamma\gamma$ through a $b$-quark loop. The final-state photons
    have light-like momenta $k_1$ and $k_2$. 
We define two light-like vectors, $n_{1}$ and $n_{2}$, that are
    collinear to $k_1$ and $k_2$, respectively. They satisfy the conditions
%---------------
\begin{eqnarray}
%---------------
n_1^2=0, \qquad n_2^2=0, \qquad n_{1}\cdot n_{2}=2.
%---------------
\end{eqnarray}
%---------------
Then, any four vector $\ell$ can be decomposed as 
%---------------
\begin{eqnarray}
%---------------
\ell^{\mu} 
= 
(n_{1}\cdot \ell)\frac{n_{2}^{\mu}}{2} 
+ (n_{2}\cdot \ell)\frac{n_{1}^{\mu}}{2} 
+ \ell_{\perp}^{\mu} 
= 
\ell_{+}\frac{n_{2}^{\mu}}{2} 
+ \ell_{-}\frac{n_{1}^{\mu}}{2} 
+ \ell_{\perp}^{\mu},
%---------------
\end{eqnarray}
%---------------
where $\ell_{\perp}^{\mu} = (0,\ell^{1},\ell^{2},0)$ and we have defined 
%---------------
\begin{eqnarray}
%---------------
\ell_{+}\equiv n_{1}\cdot \ell,  
\quad
\ell_{-} \equiv n_{2}\cdot \ell.
%---------------
\end{eqnarray}
%---------------
%It is customary to represent a momentum in these light-cone coordinates by 
%\beqn
%\ell^{\mu} = (\ell_{+},\ell_{-},\ell_{\perp}).
%\eeqn
Consequently, the scalar product of two momenta is given by
%---------------
\begin{eqnarray}
%---------------
k\cdot \ell 
&=& 
\frac{k_{+}\ell_{-}}{2}+\frac{k_{-}\ell_{+}}{2} +k_{\perp}\cdot \ell_{\perp} 
= 
\frac{k_{+}\ell_{-}}{2}+\frac{k_{-}\ell_{+}}{2}
-\bs{k}_{\perp}\cdot \bs{\ell}_{\perp},
%---------------
\end{eqnarray}
%---------------
where $\bs{\ell}_{\perp}$ and $\bs{k}_{\perp}$ are Euclidean two-vectors.

\section{Factorization theorem \label{sec:factorization-thm}}
 
A factorization formula for the amplitude for $H\to \gamma\gamma$
      through $b$-quark loop is given in Ref.~\cite{Liu:2019oav}. It
      holds up to corrections of relative order $m_b/m_H$ and can be
      written as 
\begin{eqnarray}
\label{eq:factorization-form}%
\mathcal{M}_b(H\to \gamma\gamma) =\sum_{i=1}^3 H_i \langle
\gamma\gamma|{\cal O}_i|H\rangle,
\end{eqnarray}
where the ${\cal O}_i$ are operators that are defined in
Ref.~\cite{Liu:2019oav}, the $H_i$ are hard matching coefficients
(Wilson coefficients), and the products of the $H_i$ and the operator
matrix elements are in the convolutional sense. ${\cal O}_1$ is a
contact interaction between the Higgs and photon fields.  ${\cal O}_2$
is a sum of two contributions, one involving the photon with momentum
$k_1$ and the corresponding SCET collinear field and the other involving
the photon with momentum $k_2$ and the corresponding SCET collinear
field. ${\cal O}_3$ involves hard-collinear SCET fields for both the
$k_1$ and $k_2$ directions and a soft-quark field.  Explicit
      expressions for these operator matrix elements and graphical
      representations of them are given in Ref.~\cite{Liu:2019oav}.

After applying the field redefinitions \cite{Bauer:2001yt} which
decouple the hard-collinear fields from the soft gluons, one can write
the factorized form of ${\cal O}_3$ as follows \cite{Liu:2019oav}:
%---------------
\begin{eqnarray}
\label{eq:O3-fac}
%---------------
{\cal O}_3
&=&
H(0)
\int d^Dx \int d^Dy\,
T
\bigg\{
\left[\left(\mathcal{A}\sl^\perp_{n_1}(x)+\mathcal{G}\sl^\perp_{n_1}(x)\right)
\mathcal{X}_{n_1}(x)\right]^{\alpha i}
\bar{\mathcal{X}}^{\beta j}_{n_1}(0)
\bigg\}
\nonumber \\
&&
\times
T
\bigg\{
\mathcal{X}^{\beta k}_{n_2}(0)
\left[\bar{\mathcal{X}}_{n_2}(y)
\left(\mathcal{A}\sl^\perp_{n_2}(y)+\mathcal{G}\sl^\perp_{n_2}(y)\right)
\right]^{\gamma l}
\bigg\}
\nonumber \\
&&
\times 
T
\bigg\{
\left[
S^\dagger_{n_2}(y_+)
q_s(y_+)
\right]^{\gamma l}
\left[
\bar{q}_s(x_-)
S_{n_1}(x_-)
\right]^{\alpha i}
\left[S^\dagger_{n_1}(0)S_{n_2}(0)\right]^{jk}
\bigg\}+\textrm{h.c.},
%---------------
\end{eqnarray}
%---------------
where h.c. denotes the hermitian conjugate contributions, $T$ denotes a
time-ordered product, and $(i,j,k,l)$ and $(\alpha,\beta,\gamma)$ are
color and spin indices, respectively.  $H$ is the Higgs-boson field, and
$q_s$ is the soft-quark SCET field.  $\mathcal{A}^\mu_{n_i}$,
$\mathcal{G}^\mu_{n_i}$, and $\mathcal{X}_{n_i}$ are the building blocks
of SCET for the $n_i$-hard-collinear photon, gluon, and quark fields,
respectively.  The $S_{n_i}$ are soft Wilson lines (in distinction
to the collinear Wilson lines, which are contained in the $\mathcal{X}_{n_i}$), which are defined by
%---------------
\begin{eqnarray}
%---------------
\label{def:Wilson-line}%
S_{n_{i}}(x)&=& P~ \text{exp}\left[ig_{s}\int_{-\infty}^{0}dt~ n_{i}\cdot G_{s}(x+tn_{i})\right].
%---------------
\end{eqnarray}
%---------------
Here, $G_{s}$ is the soft-gluon field.

In Eq.~(\ref{eq:O3-fac}), the first and second time-ordered products
are the jet operators that account for the contributions that
    are collinear to $n_1$ and $n_2$, respectively, and the last
time-ordered product is the soft operator. In
Ref.~\cite{Liu:2020eqe}, the authors take matrix elements of the jet
operators between the vacuum and one-photon states, take the
vacuum-to-vacuum matrix element of the soft operator, take Fourier
transforms of these matrix elements, extract some kinematic and
Dirac-matrix factors, and make use of analyticity properties of the
matrix elements to arrive at a factorized form for $\langle
\gamma\gamma|{\cal O}_3|H\rangle$ that depends only on scalar jet and
soft functions:
%---------------
\begin{eqnarray}
%---------------
\label{eq:O3}%
\langle \gamma\gamma|{\cal O}_3|H \rangle
=
2
g_\perp^{\mu\nu}
\int_0^\infty \frac{d\omega}{\omega}
S_{1}(\omega)
\int_{\sqrt\omega}^\infty
\frac{d\ell_-}{\ell_-}
J(m_H\omega/\ell_-) J(-m_H\ell_-).
%---------------
\end{eqnarray}
%---------------
Here, $\omega=\ell_+\ell_-$, $S_{1}(\omega)$ is the discontinuity of a
structure function of the soft function, which we define in
Eq.~(\ref{eq:disc-struct-fn}) below, and $J$ is the radiative jet
function \cite{DelDuca:1990gz,Bonocore:2015esa,Bonocore:2016awd} which
describes the emission of collinear gluons from a collinear $b$ quark.
The radiative jet function also appears in the radiative $B$-meson decay
$B^-\to\gamma\ell^-\bar{\nu}$
\cite{Lunghi:2002ju,Bosch:2003fc,Wang:2016qii,Wang:2018wfj}.  The
    integrations in Eq.~(\ref{eq:O3}) contain rapidity divergences as
    $\omega$ and $\ell_-$ tend to infinity and are well defined only
    after one has imposed a rapidity regulator or a cutoff
    \cite{Liu:2020eqe}.

\section{Operator definition of the soft function \label{sec:soft-op-defn}}
Following Ref.~\cite{Liu:2020eqe}, we define the soft function 
in terms of a vacuum-to-vacuum matrix element of the product of
the soft-quark propagator with soft Wilson lines:
%---------------
\begin{eqnarray}
%---------------
\label{eq:soft-function}%
\bs{S}(\ell_+,\ell_-) &=& 
\frac{2i\pi}{N_c}
\int dx_- dy_{+}  \exp\left[i\frac{\ell_{-}y_{+}-\ell_{+}x_{-}}{2}\right]\nn\\
&&\times
\langle 0|
T\,\text{Tr}\biggl[
S_{n_{2}}(0)S^{\dagger}_{n_{2}}(y_{+},0,\bs{0}_\perp)
q_{s}(y_{+},0,\bs{0}_\perp)\bar{q}_{s}(0,x_{-},\bs{0}_\perp)
S_{n_{1}}(0,x_{-},\bs{0}_\perp)S^{\dagger}_{n_{1}}(0)
\biggr]
|0\rangle.
\nn\\
%---------------
\end{eqnarray}
%---------------
Here, $N_c=3$ is the number of quark colors, and the trace is over
color, but not spinor, indices. The products of semi-infinite Wilson lines
$S_{n_i}S_{n_i}^\dagger$ in Eq.~(\ref{eq:soft-function}) can be written
as finite-length Wilson lines \cite{Liu:2020eqe}. We find it more
convenient in our calculations to keep the Wilson lines in the product
form. However, one should bear in mind that, because the product of
semi-infinite Wilson lines yields a Wilson line of finite length, the
rapidity divergences that are associated with the individual
semi-infinite Wilson lines cancel.

The expression for the soft function contains an implicit integration
over the transverse momentum of the soft quark. It is convenient to make
this integration manifest, which we accomplish by defining an
unintegrated soft function
%---------------
\begin{eqnarray}
%---------------
\bs{S}(\ell_{+},\ell_{-},\ell_\perp)
&=& 
\frac{2i\pi}{N_{c}}
\int dx_- dy_+ d^{D-2}\bs{z}_\perp
\exp
\left[i\left(\frac{\ell_{-}y_{+}-\ell_{+}x_{-}}{2}-\bs{\ell}_{\perp}\cdot
\bs{z}_\perp \right)\right]
\nonumber \\
&&
\times
\langle 0|T\,\text{Tr}\biggl[
S_{n_{2}}(0,0,\bs{z}_\perp/2)S^{\dagger}_{n_{2}}(y_+,0,\bs{z}_\perp/2)
q_{s}(y_+,0,\bs{z}_\perp/2)
\nonumber \\
&&
\quad~
\times \bar{q}_{s}(0,x_-,-\bs{z}_\perp/2)
S_{n_{1}}(0,x_-,-\bs{z}_\perp/2)
S^{\dagger}_{n_{1}}(0,0,-\bs{z}_\perp/2)\biggr]
|0\rangle,
%---------------
\end{eqnarray}
%---------------
where
%---------------
\begin{eqnarray}
%---------------
\int_{\bs{\ell}_{\perp}}
\bs{S}(\ell_+,\ell_-,\ell_\perp)&=&
\bs{S}(\ell_+,\ell_-),
%---------------
\end{eqnarray}
%---------------
with 
%---------------
\begin{eqnarray}
%---------------
\int_{\bs{\ell}_{\perp}} &\equiv&
\int\frac{d^{D-2}\bs{\ell}_{\perp}}{(2\pi)^{D-2}},  
%---------------
\end{eqnarray}  
%--------------- 
where the number of space-time dimensions $D=4-2\epsilon$ is used to regularize divergent 
integrals.

\section{Structure functions and discontinuities \label{sec:struct-fns}}

We wish to decompose the soft function into structure functions. The
reparametrization invariance of the soft function requires that any
numerator factor $n_1$ ($n_2$) be accompanied by a denominator
factor $n_1$ ($n_2$) \cite{Manohar:2002fd,Liu:2020eqe}. (We use
$n_1\cdot n_2=2$ to eliminate factors $n_1\cdot n_2$.)
Then we can decompose the unintegrated soft function
$\bs{S}(\ell_{+},\ell_{-},{\ell}_{\perp})$ as follows:
%---------------
\begin{eqnarray}
\label{eq:sf}
%---------------
\bs{S}(\ell_+,\ell_-,\ell_\perp)
&=&
m_b\bs{S}_{1}(\omega,\bs{\ell}_\perp^2)
+
\frac{\slashed{n}_{1}}{2}(n_{2}\cdot \ell)
\bs{S}_{2}(\omega,\bs{\ell}_\perp^2)
+
\frac{\slashed{n}_{2}}{2}(n_{1}\cdot \ell)
\bs{S}_{3}(\omega,\bs{\ell}_\perp^2)
+
m_b\frac{\slashed{n}_{2}\slashed{n}_{1}}{4}
\bs{S}_{4}(\omega,\bs{\ell}_\perp^2)
\nonumber \\
&&
+
\slashed{\ell}_{\perp}
\bs{S}_{5}(\omega,\bs{\ell}_\perp^2)
+
\frac{m_b\slashed{n}_{1}\slashed{\ell}_{\perp}}{2(n_{1}\cdot \ell)}
\bs{S}_{6}(\omega,\bs{\ell}_\perp^2)
+
\frac{m_b\slashed{\ell}_{\perp}\slashed{n}_{2}}{2(n_{2}\cdot \ell)}
\bs{S}_{7}(\omega,\bs{\ell}_\perp^2)
+
\frac{\slashed{n}_{2}\slashed{\ell}_{\perp}\slashed{n}_{1}}{4}
\bs{S}_{8}(\omega,\bs{\ell}_\perp^2),
\nonumber \\
%---------------
\end{eqnarray}
%---------------
where the structure functions $\bs{S}_i$ are scalar-valued functions of
$\omega=\ell_+\ell_-$ and $\bm{\ell}_\perp^2$. The Dirac structures are
the most general parity-even ones that can be obtained from the
four-vectors $\ell$, $n_1$, and $n_2$, subject to the
reparametrization-invariance constraints. In selecting this particular
decomposition into linearly independent Dirac structures, we have
observed the convention that $\slashed{n}_1$ always appears to the right
of $\slashed{n}_2$. This will prove to be convenient when we take into
    account
the Dirac structure of the jet and hard factors in the factorization
theorem.  The decomposition of the integrated soft function into
structure functions is given by
%---------------
\begin{eqnarray}
\label{eq:sff}
%---------------
\bs{S}(\ell_+,\ell_-)
&=&
m_{b}\bs{S}_{1}(\omega)
+
\frac{\slashed{n}_{1}}{2}(n_{2}\cdot \ell)
\bs{S}_{2}(\omega)
+
\frac{\slashed{n}_{2}}{2}(n_{1}\cdot \ell)
\bs{S}_{3}(\omega)
+
m_{b}\frac{\slashed{n}_{2}\slashed{n}_{1}}{4}
\bs{S}_{4}(\omega),
%---------------
\end{eqnarray}
%---------------
where the integrated form factors are defined by
%---------------
\begin{eqnarray}
\label{eq:integrated-soft-function-form-factor}
%---------------
\bm{S}_i(\omega)
=
\int_{\bm{\ell}_\perp}\bm{S}_{i}(\omega,\bm{\ell}_\perp^2),
\quad \textrm{for $i=$1, 2, 3, and 4.}
%---------------
\end{eqnarray}
%---------------
Note that the $\ell_\perp$-dependent contributions are now absent
in the decomposition of the integrated soft function.

It is shown in Ref.~\cite{Liu:2020eqe} that, because of the analytic
    properties of the soft function (see Sec.~\ref{sec:analyticity}) and
    the jet functions, the factorization theorem for $H\to \gamma\gamma$
    through a $b$-quark loop can be written in terms of the
    discontinuity of the soft function, which is given by
\begin{eqnarray}
\label{eq:disc}%
S(\omega)&=&\frac{1}{2\pi i}[\bs{S}(\omega+i\ve)-\bs{S}(\omega-i\ve)].
\end{eqnarray}
We use a non-bold $S$ to distinguish the discontinuity of the soft
function from the soft function $\bs{S}$. Similarly, we use $S_i$ to
denote the discontinuities of the soft structure functions $\bs{S}_i$:
\begin{eqnarray}
\label{eq:disc-struct-fn}%
S_i(\omega)&=&\frac{1}{2\pi i}[\bs{S}_i(\omega+i\ve)-\bs{S}_i(\omega-i\ve)].
\end{eqnarray}

\section{Diagrammatic form of the soft function 
\label{sec:diagrammatic-soft-fn}} 
\begin{figure}
\begin{center}
\includegraphics[width=0.33\columnwidth]{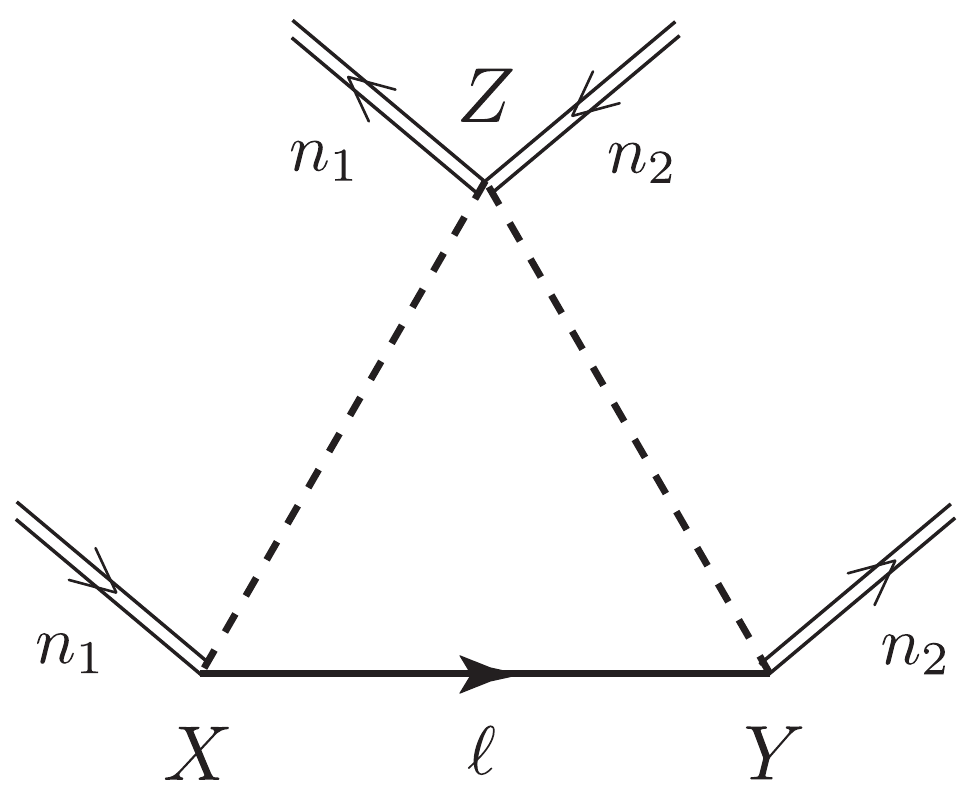}
\caption{Diagrammatic form of the integrated soft function
      $\bs{S}(\ell_+,\ell_-)$. The solid line is a
      soft-quark propagator. The double solid lines with incoming arrows
      are Wilson lines $S_{n_i}$, and the double solid lines with
      outgoing arrows are hermitian-conjugate Wilson lines
      $S_{n_i}^\dagger$. The dashed lines indicate a space-time
      separation, as is described in the text.
\label{fig:WardIdentity}
}
\end{center}
\end{figure}
\begin{figure}
\begin{center}
\includegraphics[width=0.55\columnwidth]{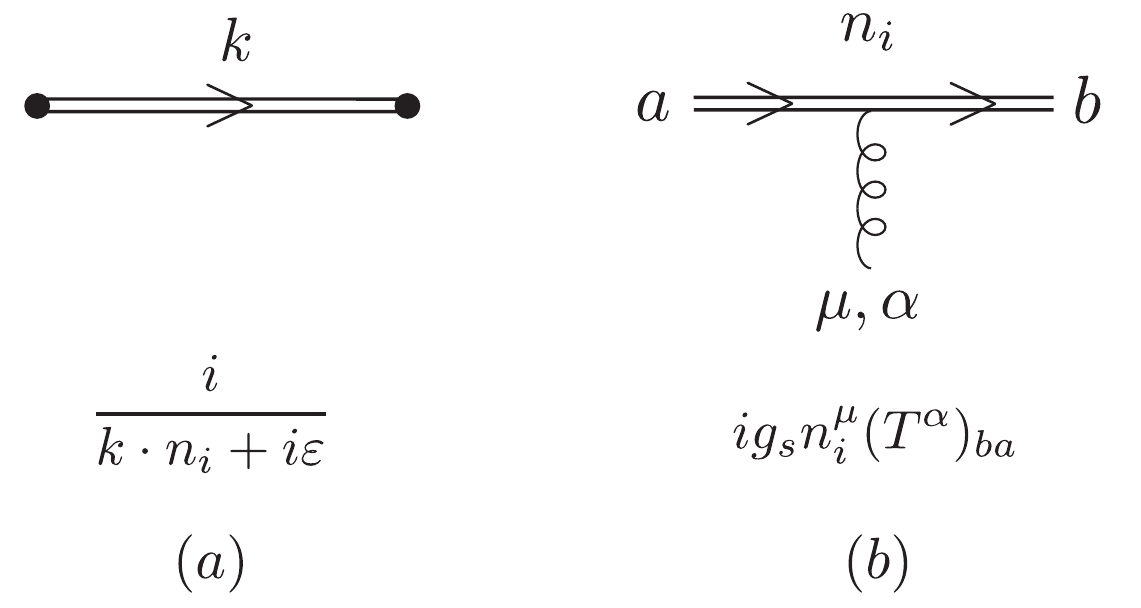}
\caption{ Feynman rules for a Wilson line $S_{n_i}$ that is collinear to
  $n_i$.  (a) The Wilson-line propagator.  (b) The vertex for the
  interaction of a gluon with the Wilson line.  $a$ and $b$ are color
  indices and $T^\alpha$ is the SU(3) generating matrix for the
  fundamental representation.  These rules apply to a diagrammatic
  Wilson line for which the arrow is incoming to a vertex. A
  diagrammatic Wilson line for which the arrow is outgoing from a vertex
  denotes a hermitian-conjugate Wilson line $S_{n_i}^\dagger$.  One obtains
  the Feynman rules for $S_{n_i}^\dagger$ from the rules in the figure by
  taking the hermitian conjugate.
\label{fig:Wilson-Feynrules}
}
\end{center}
\end{figure}

The diagrammatic form of the integrated soft function
$\bs{S}(\ell_+,\ell_-)$ is shown in Fig.~\ref{fig:WardIdentity}. For
clarity, we have suppressed gluons, which attach in all possible ways to
the Wilson lines and the soft-quark propagator. The gluons interact with
themselves, Wilson lines, soft-quarks, light-quarks, and ghosts
according to the standard rules of QCD.  The
Feynman rules for a Wilson line that is collinear to $n_i$ follow from
the definition in Eq.~(\ref{def:Wilson-line}). They are given in
Fig.~\ref{fig:Wilson-Feynrules}.

The dashed lines in Fig.~\ref{fig:WardIdentity} are not propagators but,
rather, indicate a space-time separation. There is no separation in the
transverse direction between the points $X$ and $Z$ and between the
points $Y$ and $Z$. Hence, transverse momenta can be routed between the
points $X$ and $Z$ and between the points $Y$ and $Z$. The points $X$
and $Z$ are separated in the $-$ light-front direction, but not in the
$+$ light-front direction. [That is, $n_2\cdot (X-Z)=(X-Z)_-\neq 0$, but
  $n_1\cdot (X-Z)=(X-Z)_+ =0$.]  Similarly, the points $Y$ and $Z$ are
separated in the $+$ light-front direction, but not in the $-$
light-front direction.  Hence, $-$ components of momentum can be routed
between $X$ and $Z$, and $+$ components of momentum can be routed
between $Y$ and $Z$.  The external momentum $\ell_+$ enters the diagram
at $X$, proceeds through the soft-quark propagator to $Y$, and then
proceeds to $Z$, which is a sink. Similarly, the external momentum
$-\ell_-$ enters the diagram at $Y$, proceeds through the soft-quark
propagator to $X$, and then proceeds to $Z$.  The internal momentum
$\ell_\perp$ runs in a loop from $X$ to $Y$ to $Z$ to $X$.

It is easy to understand the form of the soft function in terms of a
diagrammatic analysis in QCD. In the factorization formula in
Eq.~(\ref{eq:factorization-form}), the pinch-singular contributions to
the operator matrix element $\langle \gamma\gamma|{\cal O}_3|H\rangle$
come from a region of momentum space in which the left-hand quark line
in Fig.~\ref{fig:H2gamma} and associated gluons comprise a jet in which
all of the particles are collinear to $k_1$, the right-hand quark line
in Fig.~\ref{fig:H2gamma} and associated gluons comprise a jet in which
all of the particles are collinear to $k_2$, and the lower quark line
and associated gluons form a subgraph in which all of particles are
soft. Gluons from the soft subgraph can attach to particles in either of
the jets.

One can follow standard procedures to factor these gluons topologically
from the jets. (See, for example, Ref.~\cite{Collins:1989gx}.) First,
one makes the appropriate soft approximation (Grammer-Yennie
approximation \cite{Grammer:1973db}) for the soft-gluon attachments to
each jet. Then one applies graphical Ward identities to factor the gluon
attachments. This produces a Wilson line
$S_{n_1}$ at the lower end of the $k_1$ jet, a Wilson line
$S_{n_1}^\dagger$ at the upper end of the $k_1$ jet, a Wilson line
$S_{n_2}^\dagger $ at the lower end of the $k_2$ jet, and a Wilson line
$S_{n_2}$ at the upper end of the $k_2$ jet. The Wilson lines at the
upper ends of the jets still appear to entangle the soft gluons with the
jets. However, that entanglement can be removed by making use of the
facts that the $k_1$ jet is sensitive only to the $+$ components of
momenta that are routed through it and the $k_2$ jet is sensitive only
to the $-$ components of momenta that are routed through it. Then, one
can route the $+$ components of the momenta of gluons that attach to
$S_{n_1}^\dagger$ through the $k_2$ jet and route the $-$ components of
the momenta of gluons that attach to $S_{n_2}$ through the $k_1$
jet, thereby rendering the jet functions insensitive to the gluon
momenta in the upper Wilson lines. This unorthodox momentum routing
results in the factorization of the soft function from the jet functions
and leads to the space-time picture in Fig.~\ref{fig:WardIdentity}.

As we will see, the unorthodox flow of momenta in the soft function
    results in a nonlocal UV renormalization of the soft function.

\section{LO and NLO contributions to the soft
function \label{sec:LO-NLO-soft-fn}}

The LO soft function is given by the integral of the soft-quark
propagator over $\ell_\perp$:
%---------------
\begin{eqnarray}
%---------------
\bs{S}^{\text{LO}}(\ell_{+},\ell_{-})
&=&
m_b\bs{S}^{\text{LO}}_{1}(\omega)
+ \frac{\slashed{n}_{1}}{2}\,n_2\cdot\ell\,\bs{S}^{\text{LO}}_{2}(\omega) 
+\frac{\slashed{n}_{2}}{2}\,n_1\cdot\ell\,\bs{S}^{\text{LO}}_{3}(\omega),
%---------------
\end{eqnarray}
%---------------
where the LO structure functions are
%---------------
\begin{eqnarray}
\label{eq:SLO-exp}
%---------------
\bs{S}^{\text{LO}}_{1,2,3}(\omega)&=&
(4\pi)^\epsilon
\Gamma(\epsilon)
\left(-\omega+m_b^{2}-i\ve\right)^{-\e}.
%---------------
\end{eqnarray}
%---------------
The discontinuities of the LO soft structure functions are
\begin{eqnarray}
\label{eq:LO-disc}%
S^{\text{LO}}_{1,2,3}(\omega)&=&
\frac{(4\pi)^\epsilon}
{\Gamma(1-\epsilon)}
\left(\omega-m_b^{2}\right)^{-\e}
\theta(\omega-m_b^2)
=
\theta(\omega-m_b^2)
\left[1+O(\epsilon)\right].
\end{eqnarray}

Expressions for the NLO soft function are given in Eqs.~(4.6)--(4.8) and
(B.1) of Ref.~\cite{Liu:2019oav}. We have confirmed these
expressions.\footnote{Note that the definition of the soft function in
Ref.~\cite{Liu:2019oav} is equal to $-N_c(\alpha_{b,0}/\pi)
e^{\epsilon\gamma_E}/(4\pi)^\epsilon$ times our definition.}

\section{Analyticity of the soft function\label{sec:analyticity}}
In Ref.~\cite{Liu:2020eqe}, it is stated that the soft function
$\bs{S}(\ell_+,\ell_-)$ is analytic in the complex $\omega$ plane,
except for a cut that lies just below the positive real axis and extends
from $\omega=-i\ve$ to infinity. This is somewhat surprising, as one
might expect the cut to extend from the threshold for production of the
massive bottom quark, $\omega=m_b^2$, to infinity. This is indeed
the case for the lowest-order contribution to $\bs{S}(\ell_+,\ell_-)$.
However, as we show through some specific one-loop examples in
Appendix~\ref{sec:analyticity2}, a cut does indeed appear along the
entire $\omega$ positive real axis.  The examples are presented in
terms of $x_+$-ordered light-front perturbation theory.

The analyticity of the soft function is described by the analyticity of
its structure functions. As we have seen, the structure functions are
functions of the product $\ell_+\ell_-$.  Therefore, in considering the
analyticity of the structure functions as a function of
$\omega=\ell_+\ell_-$, we can, without loss of generality, take $\ell_+
>0$. This choice simplifies the analysis of diagrams in light-front
perturbation theory (see Appendix~\ref{sec:analyticity2}) because it
insures that the $+$ momentum of the initial soft-quark line is
always positive (flowing from left to right in the light-front
diagrams).

In light-front perturbation theory, imaginary contributions arise from
vanishing energy denominators. This occurs when the initial light-front
energy $\ell_-$ is equal to the sum of the on-shell light-front energies
of the intermediate-state lines. An imaginary part can arise in some of
the light-front diagrams when $\ell_-=\omega/\ell_+=0$ because the
light-front energy of one of the on-shell intermediate states
vanishes. This can happen because, in some light-front diagrams, a
    $b$-quark line can carry an infinite light-front longitudinal
    momentum $k_+$, causing its intermediate-state light-front energy
    $(\bs{k}_\perp^2+m_b^2)/k_+$ to vanish. In an ordinary QCD
    $b$-quark-self-energy diagram, an infinite $+$ longitudinal
    momentum cannot appear in a $b$-quark line because it would result
    in a negative longitudinal momentum in one of the diagrammatic
    lines. Negative longitudinal momenta (backward-moving lines) are
    disallowed in light-front perturbation theory. However, in the soft
    function, owing to the unorthodox momentum routing or the presence
    of an $n_2$ Wilson line, a $b$-quark line in the soft function can
    carry infinite $+$ longitudinal momentum.

The property of light-front perturbation theory that all of the $+$
longitudinal momenta are positive insures that all of the light-front
intermediate-state energies are also positive. Hence, the
energy denominators can never vanish if $\ell_-<0$. For our choice
$\ell_+>0$, this implies that the structure functions have no imaginary
parts unless $\omega$ is greater than $0$.  Therefore, we conclude that
$\bs{S}(\ell_+,\ell_-)$ is analytic in the complex $\omega$ plane,
except for a cut that runs just below the real axis from $\omega=-i\ve$
to infinity.

\section{Renormalization of the soft function \label{sec:renorm-soft-fn}}

The soft-function structure functions are renormalized as
\begin{eqnarray}
\label{eq:soft-renorm}%
\bm{S}_i^\textrm{R}(\omega)
=\sum_{j=1} 
\int_0^\infty d\omega'
Z_S^{ij}(\omega,\omega';\mu)
\bm{S}_j(\omega'),
\end{eqnarray}
where ${\bm S}^\textrm{R}$ is the renormalized soft function and $\mu$ is the
renormalization scale. As usual, $Z_S$ has an
expansion in powers of the strong coupling $\alpha_{s}$:
\begin{eqnarray}
&&Z_S=Z_{S}^{(0)}+\alpha_{s} Z_{S}^{(1)}+\alpha_{s}^2 Z_S^{(2)}+\cdots.
\end{eqnarray}

 In this paper, we compute $Z_S$ through order $\alpha_{s}$. The
 order-$\alpha_s$ contribution to $Z_S$ is the order-$\alpha_s$
 counterterm for ${\bm S}$. In minimal subtraction in dimensional
       regularization, this is the negative of the UV pole terms that
 appear in the one-loop QCD corrections to ${\bm S}$.

Note that $Z_S$ contains only the renormalizations that are associated
    with ${\bm S}$ [renormalizations of the operator in
            Eq.~(\ref{eq:soft-function})] and does not include the
    coupling-constant and mass renormalizations of QCD. It does,
    however, include the wave-function renormalization that is
    associated with the soft-quark field.

In principle, the renormalization factor $Z_S$ includes a UV divergence
that arises from the $\ell_\perp$ integration that is implicit in the
definition of the soft operator. This divergence starts at order
$\alpha_{s}^0$. [See Eq.~(\ref{eq:SLO-exp}).] We do not include this
divergence in our computations of $Z_S$ because, ultimately, we are
interested in the discontinuity of the soft function $S(\ell_+,\ell_-)$
[Eq.~(\ref{eq:disc})]. In the discontinuity of the soft function, the
$\ell_\perp$ integration does not produce a UV divergence because the
discontinuity of the soft function has support over only a finite
    range of $\bs{\ell}_\perp^2$. The finiteness of the discontinuity
of the LO soft function can be seen explicitly in
Eq.~(\ref{eq:LO-disc}).

In our calculations of the one-loop UV divergences in the soft
    function we include a factor of the all-orders soft function, along
    with the divergent loop. That is, we compute the one-loop
    counterterm corrections to the all-orders soft function. In using
    this method, it is essential to keep in mind that the all-orders
    soft function is factored from the one-loop contribution. That is,
    the loop momenta that are internal to the all-orders soft function
    do not enter into the one-loop expressions.

This method is advantageous for several reasons. First, as we have
    mentioned, the UV divergences in the soft function are nonlocal, in
    the sense that they involve integrations over the external
    longitudinal momenta of the soft function, rather than simple
    multiplications. We use the all-orders soft function to keep track
    of these integrations. Furthermore, as we will see, the UV
    divergences involve both left- and right-multiplication of the soft
    function by Dirac matrices. We use the all-orders soft function to
    keep track of these multiplications, as well.  In addition, the
    explicit presence of the all-orders soft function allows us to make
    use of its analyticity properties to simplify the forms of the
    one-loop divergences.

\section{One-loop renormalization and evolution of the soft function 
\label{sec:one-loop-calc}}

In this section, we compute the various contributions to the one-loop
renormalization of the soft function. We denote the one-loop
    corrections, including a factor of the all-orders soft function, by
    $\bs{S}_{(i)}(\ell_+,\ell_-)$, where the subscript $i=A_1, A_2, A_3,
    A_4, B,C$ denotes the class of one-loop diagram.

\subsection{Diagrams $\textbf{\textit{A}}_{\textbf{\textit{i}}}$}

\begin{figure}
\begin{center}
\includegraphics[width=0.55\columnwidth]{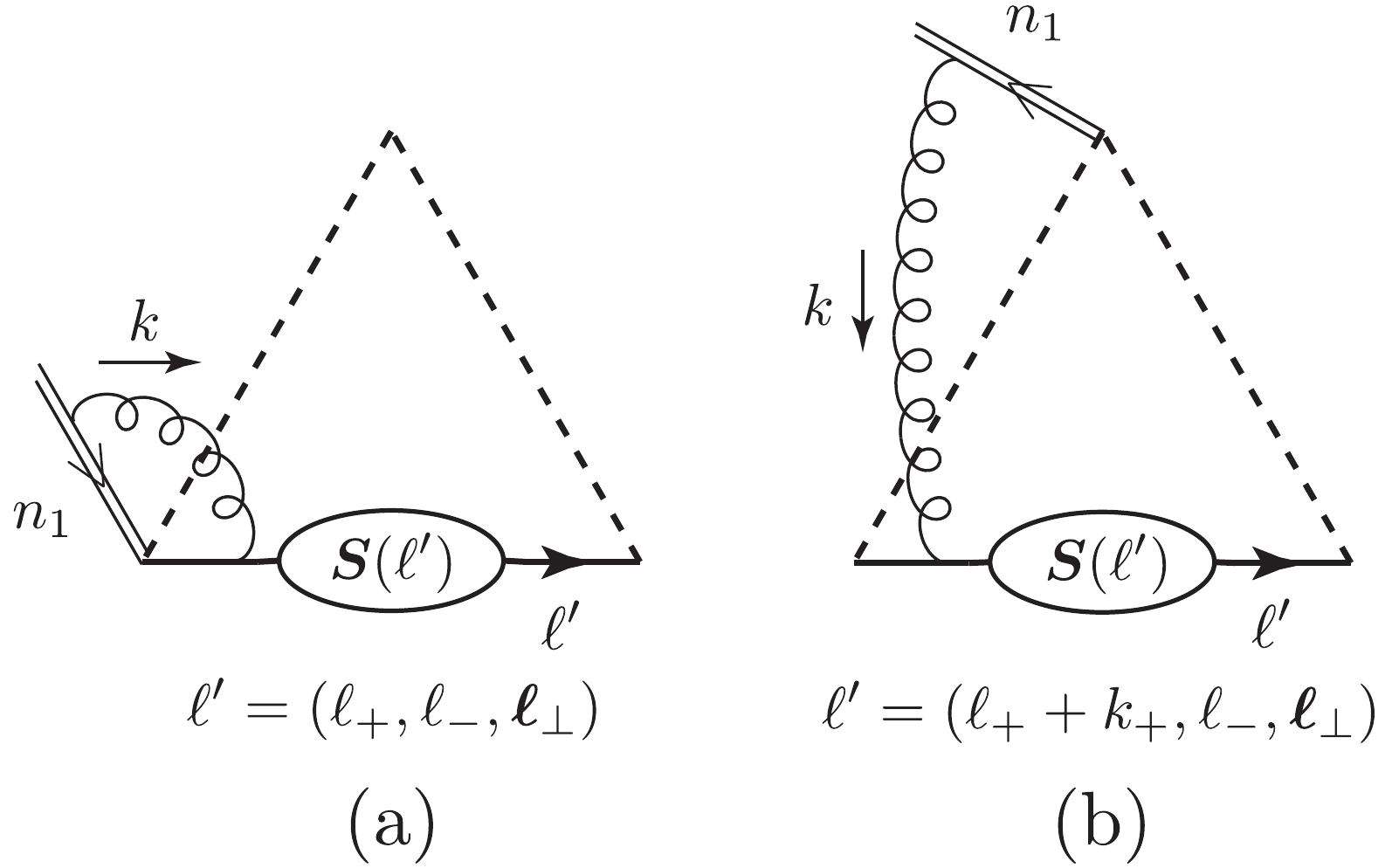}
\caption{The diagrams that contribute to $\bs{S}_{(A_1)}$. The
      blob represents the all-orders soft function. Its Wilson lines and
      the connections of gluons to them are suppressed.
\label{fig:diagrama}
}
\end{center}
\end{figure}

The expression that corresponds to the diagrams that are shown in 
Fig.~\ref{fig:diagrama} is
\begin{eqnarray}
\label{eq:figurea-starting-point}%
%---------------
\bs{S}_{(A_1)}(\ell_+,\ell_-)
&=&
ig_s^{2}C_F\left(\frac{\mu^{2} e^{\gamma_{E}}}{4\pi}\right)^{\epsilon}
\int_{\bs{\ell}_{\perp}}\int \frac{dk_+ dk_-d^{D-2}k_\perp}{(2\pi)^D}
\frac{1}
{(-k_++i\varepsilon)
(k^2+i\varepsilon)}\nonumber\\
&&\times\Bigg\{
-\frac{\bs{S}(\ell_+,\ell_-,\ell_{\perp})\frac{\slashed{n}_{1}}{2}\left[\frac{\slashed{n}_{2}}{2}\left(\ell_{+}-k_{+}\right)+\slashed{\ell}_{\perp}-\slashed{k}_{\perp}+m_b\right]}{(\ell_{+}-k_{+})(\ell_{-}-k_{-})-(\bs{\ell}_\perp-\bs{k}_\perp)^{2}-m_b^{2}+i\varepsilon}
\nn\\
&&
\quad~\,\,
+ \frac{\bs{S}(\ell_++k_{+},\ell_-,\ell_{\perp})\frac{\slashed{n}_{1}}{2}(\frac{\slashed{n}_{2}}{2}\ell_{+}+\slashed{\ell}_{\perp}-\slashed{k}_{\perp}+m_b)}{\ell_{+}(\ell_{-}-k_{-})-(\bs{\ell}_\perp-\bs{k}_\perp)^{2}-m_b^{2}+i\varepsilon}
\Bigg\},
%---------------
\end{eqnarray}
%---------------
where the first term corresponds to the diagram in
Fig.~\ref{fig:diagrama}(a) and the second term corresponds to the
diagram in Fig.~\ref{fig:diagrama}(b). We have also made the
integration over $k_\perp$ in the integrated soft function explicit.  We
note that there is a contribution that depends on
$\bs{S}(\ell_++k_+,\ell_-,\ell_\perp)$, rather than on
$\bs{S}(\ell_+,\ell_-,\ell_\perp)$. This contribution arises because of
the unorthodox momentum routing in the soft function, which is a
consequence of the form of the factorization of the soft function from
the jet functions.  As we will see, such ``nonlocal'' contributions are
a general feature of the renormalization of the soft function.

Let us initially consider the case $\ell_{+}>0$, $\ell_{-}<0$.  We
perform the $k_{-}$ integration by closing the contour in
the lower half-plane and picking up the pole at $k_{-} =
\frac{k_{\perp}^{2}}{k_{+}} -i\varepsilon$. This leads to
%---------------
\begin{eqnarray}
%---------------
\bs{S}_{(A_1)}(\ell_+,\ell_-)
&=&
-2\alpha_sC_F\left(\frac{\mu^{2} e^{\gamma_{E}}}{4\pi}\right)^{\epsilon}
\int_{\bs{\ell}_{\perp}}\int \frac{d^{D-2}k_\perp}{(2\pi)^{D-2}}\int_{0}^{\infty}\frac{dk_{+}}
{k_{+}^{2}}\nn\\
&&\times
\Bigg\{
-\frac{\theta(\ell_{+}-k_{+})\bs{S}(\ell_+,\ell_-,\ell_\perp)\frac{\slashed{n}_{1}}{2}\left[\frac{\slashed{n}_{2}}{2}\left(\ell_{+}-k_{+}\right)+\slashed{\ell}_{\perp}-\slashed{k}_{\perp}+m_b\right]}{(\ell_{+}-k_{+})\ell_{-}-\frac{\ell_{+}-{k_{+}}}{k_{+}}\bs{k}_{\perp}^{2}-(\bs{\ell}_\perp-\bs{k}_\perp)^{2}-m_b^{2}+i\varepsilon}
\nonumber\\
&&
\quad~
+
\frac{\bs{S}(\ell_++k_{+},\ell_-,\ell_{\perp})\frac{\slashed{n}_{1}}{2}(\frac{\slashed{n}_{2}}{2}\ell_{+}+\slashed{\ell}_{\perp}-\slashed{k}_{\perp}+m_b)}{\ell_{+}\ell_{-}-\frac{\ell_{+}}{k_{+}}\bs{k}_{\perp}^{2}-(\bs{\ell}_\perp-\bs{k}_\perp)^{2}-m_b^{2}+i\varepsilon}
\Bigg\}
.
%---------------
\end{eqnarray}
%---------------
Translating the integration variable $k_\perp$ according to
$k_{\perp}\rightarrow
k_{\perp}+\frac{k_{+}}{\ell_{+}+k_{+}}\ell_{\perp}$ and
$k_{\perp}\rightarrow k_{\perp}+\frac{k_{+}}{\ell_{+}}\ell_{\perp}$
for the first and second terms in the integrand, respectively, we
obtain
%---------------
\begin{eqnarray}
%---------------
\label{eq:one-w-line-integrated}%
\bs{S}_{(A_1)}(\ell_+,\ell_-)
&=&
2\alpha_sC_F\left(\frac{\mu^{2} e^{\gamma_{E}}}{4\pi}\right)^{\epsilon}
\int_{\bs{\ell}_{\perp}}\int \frac{d^{D-2}k_\perp}{(2\pi)^{D-2}}\int_{0}^{\infty}\frac{dk_{+}}
{k_{+}}\nn\\
&&\times
\Bigg\{
-\frac{\theta(\ell_{+}-k_{+})}{\ell_{+}}
\frac{\bs{S}(\ell_+,\ell_-,\ell_\perp)\frac{\slashed{n}_{1}}{2}\left[\frac{\slashed{n}_{2}}{2}\left(\ell_{+}-k_{+}\right)+\slashed{\ell}_{\perp}\frac{\ell_{+}-k_{+}}{\ell_{+}}+m_b\right]}{\bs{k}_{\perp}^{2}-\frac{k_{+}}{\ell_{+}}\left(\ell^{2}-m^{2}_{b}-\ell_{-}k_{+}+\frac{k_{+}}{\ell_{+}}\bs{\ell}_{\perp}^{2}\right)-i\varepsilon}
\nonumber\\
&&
\quad~
+
\frac{1}{\ell_{+}+k_{+}}
\frac{\bs{S}(\ell_++k_{+},\ell_-,\ell_\perp)\frac{\slashed{n}_{1}}{2}(\frac{\slashed{n}_{2}}{2}\ell_{+}+\slashed{\ell}_{\perp}\frac{\ell_{+}}{\ell_{+}+k_{+}}+m_b)}{\bs{k}_{\perp}^{2}-\frac{k_{+}\ell_{+}\left(\ell^{2}-m_b^{2}\right)+k_{+}^{2}\left(\ell_{+}\ell_{-}-m_b^{2}\right)}{\left(\ell_{+}+k_{+}\right)^{2}}-i\varepsilon}
\Bigg\}.
%---------------
\end{eqnarray}
%---------------
We note that the contributions of the individual diagrams in
Figs.~4(a) and 4(b) to Eq.~(\ref{eq:one-w-line-integrated}) are not
well defined in dimensional regularization because they contain
rapidity divergences that appear as $k_+\to 0$ with $\bs{k}_\perp^2$
fixed. However, these rapidity divergences cancel in the complete
expression in Eq.~(\ref{eq:one-w-line-integrated}). As we have
mentioned, this is as expected, since the upper semi-infinite
Wilson line cancels against the lower semi-infinite Wilson line to
produce a finite-length Wilson line. Then, only the $\bs{k}_\perp$
integration is divergent, and it produces only a UV divergence.  The
UV-divergent part is
%---------------
\begin{eqnarray}
%---------------
\bs{S}_{(A_1)}^{\text{UV}}
(\ell_+,\ell_-)
&=&
\frac{\alpha_sC_F}{2\pi}\frac{1}
{\epsilon_{\text{UV}}}\int_{\bs{\ell}_{\perp}}\int_{0}^{\infty}\frac{dk_{+}}{k_{+}}
\nonumber \\
&&
\times 
\Bigg\{
-\bs{S}(\ell_+,\ell_-,\ell_{\perp})\frac{\slashed{n}_{1}}{2}\frac{\theta(\ell_{+}-k_{+})\left[\frac{\slashed{n}_{2}}{2}\left(\ell_{+}-k_{+}\right)+\slashed{\ell}_{\perp}\frac{\ell_{+}-k_{+}}{\ell_{+}}+m_b\right]}{\ell_{+}}
\nn\\
&&
\quad~~
+
\bs{S}(\ell_++k_{+},\ell_-,\ell_{\perp})\frac{\slashed{n}_{1}}{2}\frac{(\frac{\slashed{n}_{2}}{2}\ell_{+}+\slashed{\ell}_{\perp}\frac{\ell_{+}}{\ell_{+}+k_{+}}+m_b)}{\ell_{+}+k_{+}}
\Bigg\}.
%---------------
\end{eqnarray}
%---------------
The change of variables $k_{+}= x\ell_{+}$ leads to
%---------------
\begin{eqnarray}
\label{eq:SaUV}
%---------------
\bs{S}_{(A_1)}^{\text{UV}}(\ell_+,\ell_-)
&=&
\frac{\alpha_sC_F}{2\pi}\frac{1}
{\epsilon_{\text{UV}}}\int_{\bs{\ell}_{\perp}}
\Bigg\{\bs{S}(\ell_+,\ell_-,\ell_{\perp})\frac{\slashed{n}_{1}}{2}\left(\frac{\slashed{n}_{2}}{2}+\frac{\slashed{\ell}_{\perp}}{\ell_{+}}\right)\nn\\
&&
\quad\quad\quad\quad\quad\quad~
+\int_{0}^{\infty}dx\bigg[\frac{\bs{S}(\ell_{+}(1+x),\ell_-,\ell_{\perp})}{x(1+x)}\frac{\slashed{n}_{1}}{2}\left(\frac{\slashed{n}_{2}}{2}+\frac{\slashed{\ell}_{\perp}}{\ell_{+}}\frac{1}{1+x}+\frac{m_b}{\ell_{+}}\right)
\nn\\
&&
\quad\quad\quad\quad\quad\quad\quad\quad\quad\quad\quad
-\frac{\theta(1-x)\bs{S}(\ell_+,\ell_-,\ell_{\perp})}{x}\frac{\slashed{n}_{1}}{2}\left(\frac{\slashed{n}_{2}}{2}+\frac{\slashed{\ell}_{\perp}}{\ell_{+}}+\frac{m_b}{\ell_{+}}\right)\bigg]\Bigg\}.\nn\\
%---------------
\end{eqnarray}
%---------------
In order to combine the contributions in the $x$ integration, we make
    the variable transformations $\frac{x}{1-x}\rightarrow u\rightarrow
    x$ for the terms that are proportional to
    $\bs{S}(\ell_{+},\ell_{-},\ell_{\perp})$. In doing this, we
temporarily replace the lower limit of the integration over $x$ above
with $\delta$, so that we can manipulate the two terms in the integrand
separately. We ultimately take the limit $\delta\to 0$. This procedure
    yields
%---------------
\begin{eqnarray}
\label{eq:diagramaUV}
%---------------
&&
\bs{S}_{(A_1)}^{\text{UV}}(\ell_{+},\ell_{-}) 
\nonumber \\
&=&
\frac{\alpha_s C_F}{2\pi}
\frac{1}{\epsilon_\textrm{UV}}
\int_{\bs{\ell}_{\perp}}
\Bigg[
\bm{S}(\ell_+,\ell_-,\ell_\perp)
\left(1
-\frac{n\sl_2n\sl_1}{4}
+\frac{n_2\cdot \ell}{\omega}\frac{n\sl_1\ell\sl_\perp}{2}\right)
\nonumber \\
&&
\quad\quad\quad\quad\quad\quad
+
\int_0^\infty
dx
\bigg\{
\frac{\bm{S}(\ell_+(1+x),\ell_-,\ell_\perp)}
{x(1+x)}
\left[1
-\frac{n\sl_2n\sl_1}{4}+
\frac{n_2\cdot \ell}{(1+x)\omega}\frac{n\sl_1\ell\sl_\perp}{2}
+\frac{m_b(n_2\cdot \ell)}{\omega}\frac{n\sl_1}{2}\right]
\nonumber \\
&&
\quad\quad\quad\quad\quad\quad\quad\quad\quad\quad\quad
-
\frac{\bm{S}(\ell_+,\ell_-,\ell_\perp)}
{x(1+x)}
\left[
1
-\frac{n\sl_2n\sl_1}{4}
+\frac{n_2\cdot \ell}{\omega}\frac{n\sl_1\ell\sl_\perp}{2}
+\frac{m_b(n_2\cdot \ell)}{\omega}\frac{n\sl_1}{2}
\right]
\bigg\}
\Bigg].
%---------------
\end{eqnarray}
%---------------
Note that, although the expression above was obtained for the case
$\ell_{+}>0$, $\ell_{-}<0$ ($\omega<0$), it has the correct analyticity
properties to be a valid analytic continuation of
$\bs{S}_{(A_1)}^{\text{UV}}(\ell_{+},\ell_{-})$ for all $\omega$.

\begin{figure}
\begin{center}
\includegraphics[width=0.55\columnwidth]{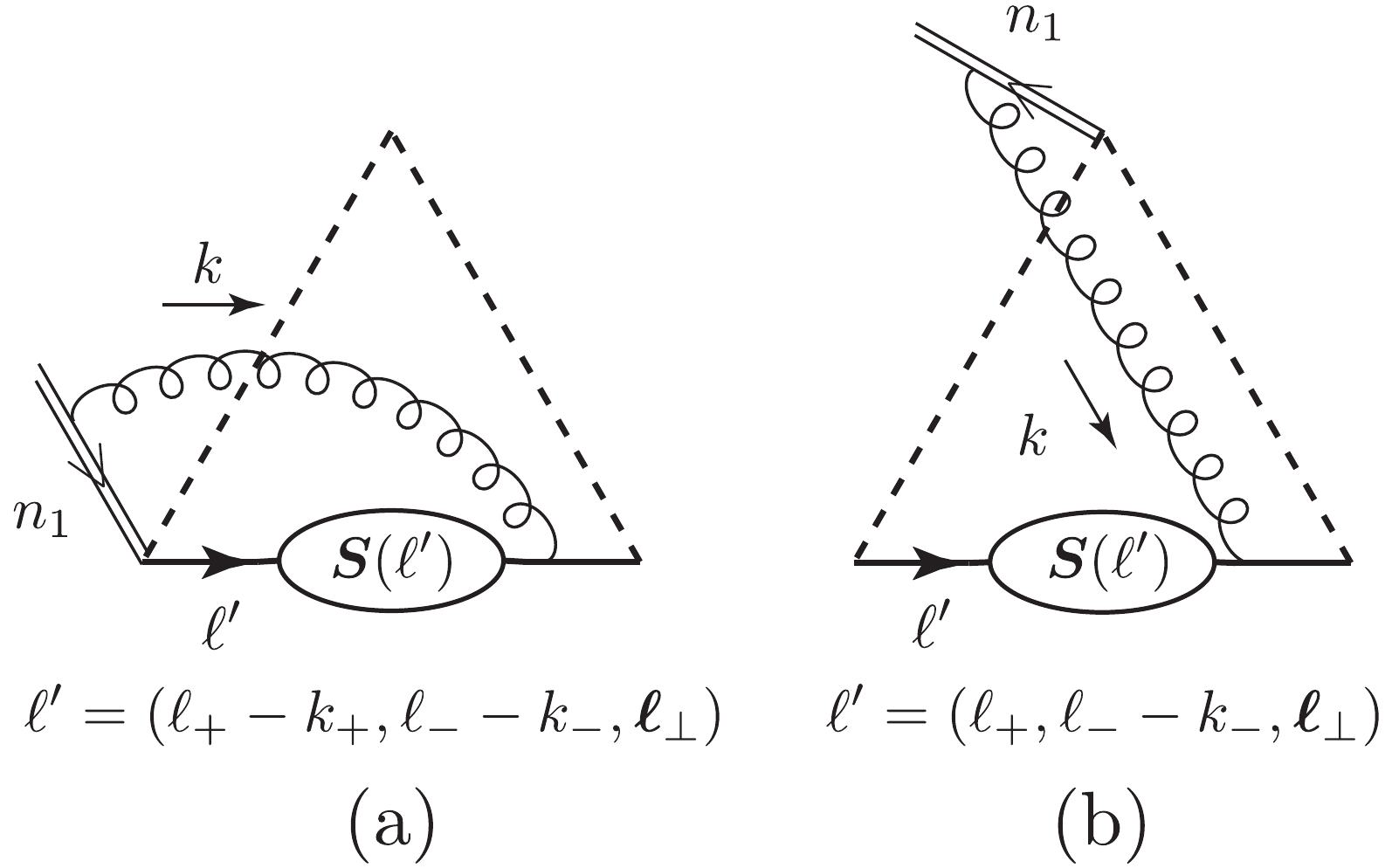}
\caption{The diagrams that contribute to $\bs{S}_{(A_2)}$
in Eq.~(\ref{eq:figurea-starting-point}).
\label{fig:diagrama2}
}
\end{center}
\end{figure}
The expression that corresponds to the diagrams that are shown in 
Fig.~\ref{fig:diagrama2} is
%---------------
\begin{eqnarray}
\label{eq:figurea2-starting-point}
%---------------
\bm{S}_{(A_2)}(\ell_{+},\ell_{-})
&=& 
i g_s^2 C_F 
\left(\frac{\mu^{2} e^{\gamma_{E}}}{4\pi}\right)^{\epsilon}
\int_{\bm{\ell}_{\perp}}
\int \frac{dk_+dk_- d^{D-2}k_\perp}{(2\pi)^D}
\frac{1}{(-k_++i\varepsilon)(k^2+i\varepsilon)}
\nonumber \\
&&
\times
\Bigg\{
-\frac{(\ell_+\frac{n\sl_2}{2}+\ell\sl_\perp+k\sl_\perp+m_b)\frac{n\sl_1}{2}
\bm{S}(\ell_{+}-k_+,\ell_{-}-k_-,\ell_\perp)}
{\ell_+\ell_--(\bm{\ell}_\perp+\bm{k}_\perp)^2-m_b^2+i\varepsilon}
\nonumber \\
&&
\quad\quad
+
\frac{\left[(\ell_++k_+)\frac{n\sl_2}{2}+\ell\sl_\perp+k\sl_\perp+m_b\right]
\frac{n\sl_1}{2}
\bm{S}(\ell_{+},\ell_{-}-k_-,\ell_\perp)}
{(\ell_++k_+)\ell_--(\bm{\ell}_\perp+\bm{k}_\perp)^2-m_b^2+i\varepsilon}
\Bigg\}.
%---------------
\end{eqnarray}
%---------------
As we have done for diagrams $A_1$, 
we consider the case $\ell_{+}>0$, $\ell_{-}<0$.
We wish to close the $k_-$ contour so as to avoid the
singularities in the functions $\bs{S}$ on the right side of
Eq.~(\ref{eq:figurea2-starting-point}). 
Let us consider the first term in the curly brackets 
of Eq.~(\ref{eq:figurea2-starting-point}).
In complex $k_-$ plane, 
the pole of $\bm{S}(\ell_{+}-k_+,\ell_{-}-k_-,\ell_\perp)$
exists in the upper half plane if $k_+<\ell_+$, and in the lower
half plane if $k_+>\ell_+$. If $k_+<0$, all of the singularities are
in the upper half plane, and the contour integration vanishes. 
Therefore, we need to consider the region $0<k_+<\ell_+$ and
close the $k_-$ contour in the lower half plane
to pick up the pole at $k_- = \frac{\bm{k}_\perp^2}{k_+}-i\varepsilon$:
%---------------
\begin{eqnarray}
%---------------
&&
\int \frac{dk_+dk_- d^{D-2}k_\perp}{(2\pi)^D}
\frac{1}{-k_++i\varepsilon}
\frac{1}{k^2+i\varepsilon}
\frac{(\ell_+\frac{n\sl_2}{2}+\ell\sl_\perp+k\sl_\perp+m_b)\frac{n\sl_1}{2}
\bm{S}(\ell_{+}-k_+,\ell_{-}-k_-,\ell_\perp)}
{\ell_+\ell_--(\bm{\ell}_\perp+\bm{k}_\perp)^2-m_b^2+i\varepsilon}
\nonumber \\
&=&
2\pi i
\int \frac{d^{D-2}k_\perp}{(2\pi)^D}
\int^{\ell_+}_{0} 
\frac{dk_+}{k_+^2}
\frac{(\ell_+\frac{n\sl_2}{2}+\ell\sl_\perp+k\sl_\perp+m_b)\frac{n\sl_1}{2}
\bm{S}(\ell_{+}-k_+,\ell_{-}-\frac{\bm{k}_\perp^2}{k_+},\ell_\perp)}
{\ell_+\ell_--(\bm{\ell}_\perp+\bm{k}_\perp)^2-m_b^2+i\varepsilon}.
%---------------
\end{eqnarray}
%---------------
Next let us consider the second term in the curly brackets 
of Eq.~(\ref{eq:figurea2-starting-point}).
The pole of $\bm{S}(\ell_{+},\ell_{-}-k_-,\ell_\perp)$
exists in the upper half plane when $\ell_+>0$.
If $k_+<0$, all of the singularities are in the upper half plane,
and the contour integration vanishes. 
Therefore, we need to consider the region $k_+>0$
and close the $k_-$ contour in the lower half plane
to pick up the pole at $k_- = \frac{\bm{k}_\perp^2}{k_+}-i\varepsilon$:
%---------------
\begin{eqnarray}
%---------------
&&
\int \frac{dk_+dk_- d^{D-2}k_\perp}{(2\pi)^D}
\frac{1}{-k_++i\varepsilon}
\frac{1}{k^2+i\varepsilon}
\frac{((\ell_++k_+)\frac{n\sl_2}{2}+\ell\sl_\perp+k\sl_\perp+m_b)
\frac{n\sl_1}{2}
\bm{S}(\ell_{+},\ell_{-}-k_-,\ell_\perp)}
{(\ell_++k_+)\ell_--(\bm{\ell}_\perp+\bm{k}_\perp)^2-m_b^2+i\varepsilon}
\nonumber \\
&=&
2\pi i
\int \frac{ d^{D-2}k_\perp}{(2\pi)^D}
\int_0^\infty 
\frac{dk_+}{k_+^2}
\frac{((\ell_++k_+)\frac{n\sl_2}{2}+\ell\sl_\perp+k\sl_\perp+m_b)
\frac{n\sl_1}{2}
\bm{S}(\ell_{+},\ell_{-}-\frac{\bm{k}_\perp^2}{k_+},\ell_\perp)}
{(\ell_++k_+)\ell_--(\bm{\ell}_\perp+\bm{k}_\perp)^2-m_b^2+i\varepsilon}.
%---------------
\end{eqnarray}
%---------------
Consequently, after the $k_-$ contour integrations, 
we can write Eq.~(\ref{eq:figurea2-starting-point}) as follows:
%---------------
\begin{eqnarray}
%---------------
\label{eq:A2-final}%
\bm{S}_{(A_2)}(\ell_{+},\ell_{-})
&=& 
2\pi g_s^2 C_F 
\left(\frac{\mu^{2} e^{\gamma_{E}}}{4\pi}\right)^{\epsilon}
\int_{\bm{\ell}_{\perp}}
\int \frac{d^{D-2}k_\perp}{(2\pi)^D}
\nonumber \\
&&
\times
\Bigg\{
\int^{\ell_+}_{0} 
\frac{dk_+}{k_+^2}
\frac{(\ell_+\frac{n\sl_2}{2}+\ell\sl_\perp+k\sl_\perp+m_b)\frac{n\sl_1}{2}
\bm{S}(\ell_{+}-k_+,\ell_{-}-\frac{\bm{k}_\perp^2}{k_+},\ell_\perp)}
{\ell_+\ell_--(\bm{\ell}_\perp+\bm{k}_\perp)^2-m_b^2+i\varepsilon}
\nonumber \\
&&
\quad~
-
\int_0^\infty 
\frac{dk_+}{k_+^2}
\frac{((\ell_++k_+)\frac{n\sl_2}{2}+\ell\sl_\perp+k\sl_\perp+m_b)
\frac{n\sl_1}{2}
\bm{S}(\ell_{+},\ell_{-}-\frac{\bm{k}_\perp^2}{k_+},\ell_\perp)}
{(\ell_++k_+)\ell_--(\bm{\ell}_\perp+\bm{k}_\perp)^2-m_b^2+i\varepsilon}
\Bigg\}.
\phantom{X}
%---------------
\end{eqnarray}
%---------------
UV divergences can potentially arise from the $\bm{k}_\perp$ or $k_+$
integrations in this expression.  In order to test for one-loop UV 
divergences, we replace the all-orders soft function with the LO
soft-quark propagator:
\begin{eqnarray}
\bm{S}(\ell_{+}-k_+,\ell_{-}-\tfrac{\bm{k}_\perp^2}{k_+},\ell_\perp)&\to&
\frac{\frac{1}{2}(\ell_+-k_+)\slashed{n}_2+
\frac{1}{2}(\ell_--\frac{\bm{k}_\perp^2}{\ell_+})\slashed{n}_1
+\slashed{\ell}_\perp+m_b}
{(\ell_+-k_+)(\ell_--\frac{\bm{k}_\perp^2}{\ell_+})-
\bm{\ell}_\perp^2-m_b^2+i\ve},\nonumber\\
\bm{S}(\ell_{+},\ell_{-}-\tfrac{\bm{k}_\perp^2}{k_+},\ell_\perp)&\to&
\frac{\frac{1}{2}\ell_+\slashed{n}_2+
\frac{1}{2}(\ell_--\frac{\bm{k}_\perp^2}{\ell_+})\slashed{n}_1
+\slashed{\ell}_\perp+m_b}
{\ell_+(\ell_--\frac{\bm{k}_\perp^2}{\ell_+})-
\bm{\ell}_\perp^2-m_b^2+i\ve}.
\end{eqnarray}
Because of the numerator factors $\slashed{n}_1$ in
Eq.~(\ref{eq:A2-final}), the terms in the propagator numerators that are
proportional to $\slashed{n}_1$ vanish. It is then easy to see that the
$\bm{k}_\perp$ and $k_+$ integrations are UV convergent. Therefore, the
diagrams $A_2$ do not contribute to the one-loop renormalization of the
soft function.

\begin{figure}
\begin{center}
\includegraphics[width=0.55\columnwidth]{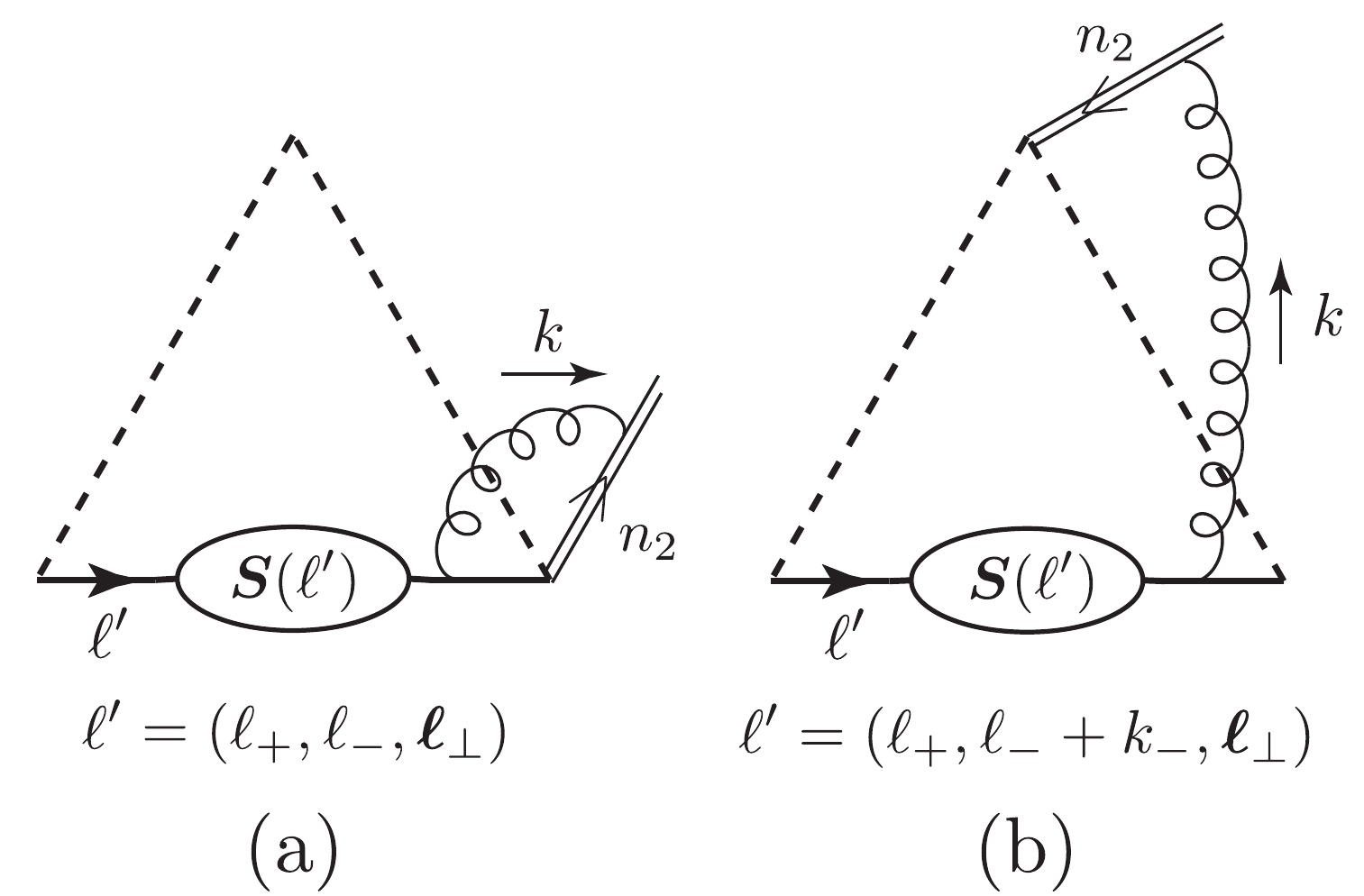}
\caption{The diagrams that contribute to $\bs{S}_{(A_3)}$
in Eq.~(\ref{eq:diagramapUV}).
\label{fig:diagramap}
}
\end{center}
\end{figure}
The mirror images of the diagrams $A_1$, which we call $A_3$, 
are shown in Fig.~\ref{fig:diagramap}. 
The expression that corresponds to these diagrams is
%---------------
\begin{eqnarray}
%---------------
\bm{S}_{(A_3)}(\ell_{+},\ell_{-},\ell_\perp)
&=&
i g_s^2 C_F
\left(\frac{\mu^2e^{\epsilon\gamma_\textrm{E}}}{4\pi}\right)^{\epsilon}
\int_{\bs{\ell}_{\perp}}
\int \frac{dk_+dk_-d^{D-2}k_\perp}{(2\pi)^D}
\frac{1}{k_-+i\varepsilon}
\frac{1}{k^2+i\varepsilon}
\nonumber \\
&&
\times
\Bigg\{
\frac{\left[
\frac{n\sl_1}{2}(\ell_- - k_-)+\ell\sl_\perp-k\sl_\perp+m_b\right]\frac{n\sl_2}{2}
\bm{S}(\ell_{+},\ell_{-},\ell_\perp)}
{(\ell_+-k_+)(\ell_--k_-)-(\bm{\ell}_\perp-\bm{k}_\perp)^2-m_b^2+i\varepsilon
}
\nonumber \\
&&
\quad\quad
-
\frac{(
\frac{n\sl_1}{2}\ell_-+\ell\sl_\perp-k\sl_\perp+m_b)\frac{n\sl_2}{2}
\bm{S}(\ell_{+},\ell_{-}+k_-,\ell_\perp)}
{(\ell_+-k_+)\ell_--(\bm{\ell}_\perp-\bm{k}_\perp)^2-m_b^2+i\varepsilon
}
\Bigg\},
%---------------
\end{eqnarray}
%---------------
where the first term corresponds to the diagram in
Fig.~\ref{fig:diagramap}(a) and the second term corresponds to the
diagram in Fig.~\ref{fig:diagramap}(b). We treat this expression along
the same lines as our treatment of the expression for the diagrams $A_1$,
except that the roles of $k_+$ and $k_-$ are interchanged and,
initially, we consider the case $\ell_+<0$, $\ell_->0$.  The result for
the UV-divergent part, valid for all $\omega$, is
%---------------
\begin{eqnarray}
\label{eq:diagramapUV}
%---------------
&&
\bs{S}^{\text{UV}}_{(A_3)}(\ell_{+},\ell_{-}) 
\nonumber \\
&=& 
\frac{\alpha_s C_F}{2\pi}
\frac{1}{\epsilon_\textrm{UV}}
\int_{\bs{\ell}_{\perp}}
\Bigg[
\left(
1-\frac{n\sl_2 n\sl_1}{4}+\frac{n_1\cdot \ell}{\omega}
\frac{\ell\sl_\perp n\sl_2}{2}\right)
\bm{S}(\ell_{+},\ell_{-},\ell_\perp)
\nonumber \\
&&
\quad\quad\quad\quad\quad\quad
+
\int_0^\infty dx 
\bigg\{
\left[
1-\frac{n\sl_2 n\sl_1}{4}+\frac{n_1\cdot \ell}{(1+x)\omega}
\frac{\ell\sl_\perp n\sl_2}{2}
+\frac{m_b(n_1\cdot \ell)}{\omega}\frac{n\sl_2}{2}\right]
\frac{
\bm{S}(\ell_{+},(1+x)\ell_{-},\ell_\perp)}
{
x(1+x)}
\nonumber \\
&&
\quad\quad\quad\quad\quad\quad\quad\quad\quad\quad\quad
-
\left[
1-\frac{n\sl_2 n\sl_1}{4}+\frac{n_1\cdot \ell}{\omega}\frac{\ell\sl_\perp n\sl_2}{2}
+\frac{m_b(n_1\cdot \ell)}{\omega}\frac{n\sl_2}{2}\right]
\frac{
\bm{S}(\ell_{+},\ell_{-},\ell_\perp)}
{x(1+x)}
\bigg\}
\Bigg].
%---------------
\end{eqnarray}
%---------------

\begin{figure}
\begin{center}
\includegraphics[width=0.55\columnwidth]{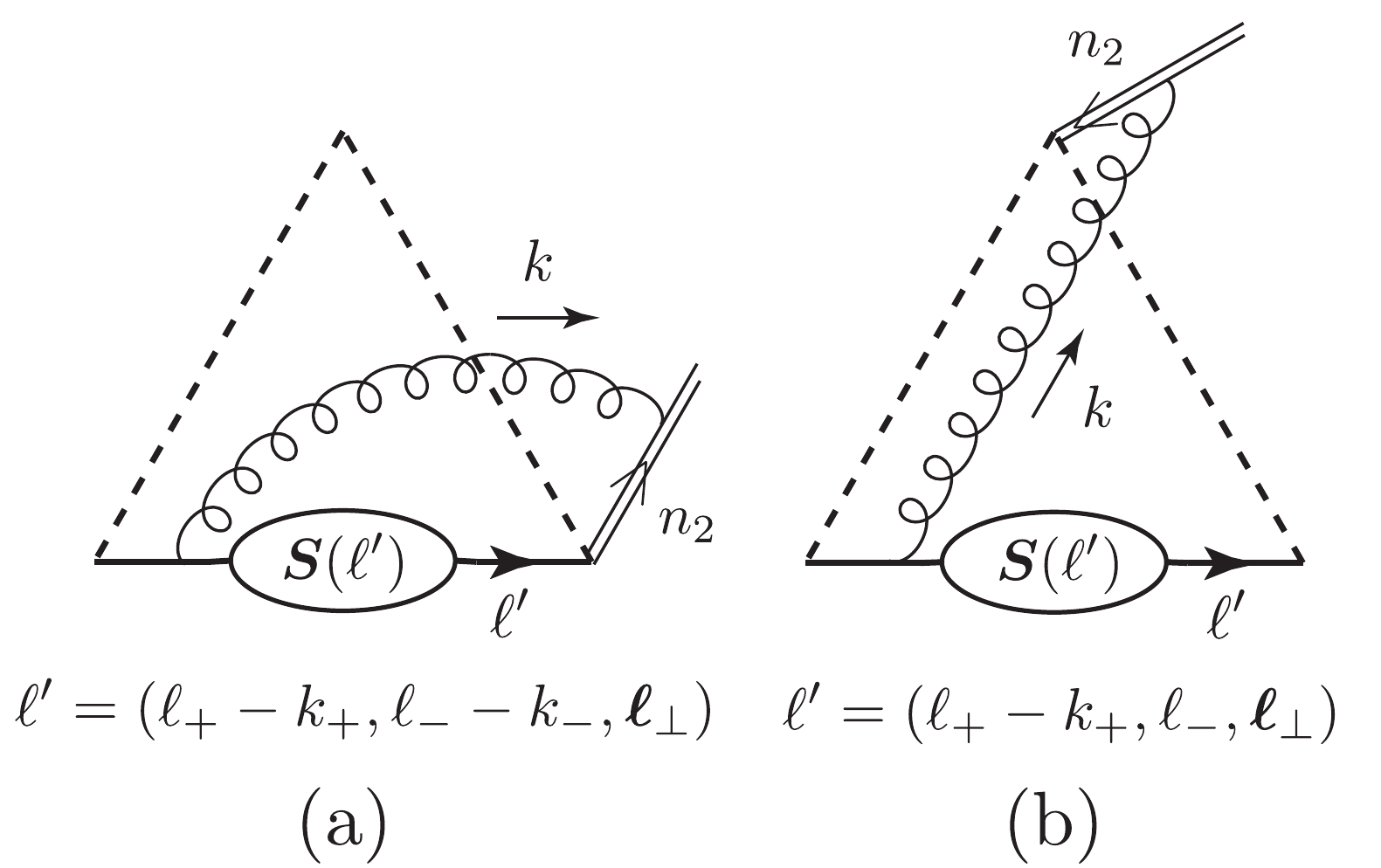}
\caption{The diagrams that contribute to $\bs{S}_{(A_4)}$
in Eq.~(\ref{eq:diagramapUV}).
\label{fig:diagrama4}
}
\end{center}
\end{figure}
The mirror images of the diagrams $A_2$, which we call $A_4$, are shown
in Fig.~\ref{fig:diagrama4}.  As with the case of diagrams $A_2$,
the diagrams $A_4$ do not contribute the UV poles, and, so, do
not contribute to the renormalization of the soft function.

\subsection{Diagrams $\textbf{\textit{B}}$}
\begin{figure}
\begin{center}
\includegraphics[width=1\columnwidth]{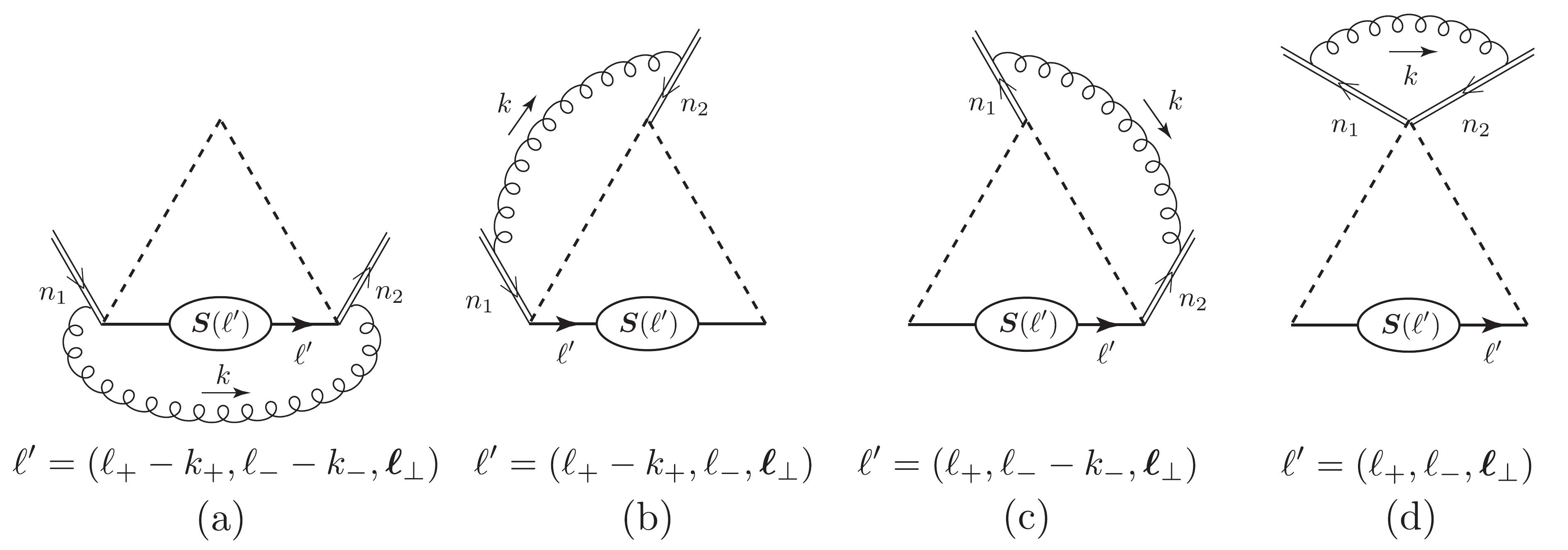}
\caption{ The diagrams that contribute to $\bs{S}_{(B)}$ in
  Eq.~(\ref{eq:S4-start}).
\label{fig:diagramc}
}
\end{center}
\end{figure}
The expression that corresponds to the diagrams that are shown in 
Fig.~\ref{fig:diagramc} is
%---------------
\begin{eqnarray}
%---------------
\label{eq:S4-start}
\bs{S}_{(B)}(\ell_+,\ell_-)
&=&
ig_s^2 C_F
\left(\frac{\mu^2e^{\gamma_\textrm{E}}}{4\pi}\right)^\epsilon
\int \frac{dk_+ dk_- d^{D-2}k_\perp}{(2\pi)^D}
\frac{1}{(-k_++i\varepsilon)(k_-+i\varepsilon)(k^2+i\varepsilon)}
\nonumber \\
&&
\times
\left[
\bm{S}(\ell_+-k_+,\ell_--k_-)
-\bm{S}(\ell_+-k_+,\ell_-)
-\bm{S}(\ell_+,\ell_--k_-)
+\bm{S}(\ell_+,\ell_-)
\right].
\phantom{XX}
%---------------
\end{eqnarray}
%---------------
Here, we have carried out the integration over $\ell_\perp$, replacing
the unintegrated soft functions on the right side of the equation with
integrated soft functions.

For the expression in Eq.~(\ref{eq:S4-start}), we initially consider
the case $\ell_{-}>0$, $\ell_{+}<0$. We choose to complete the $k_-$
integration first. We wish to close the $k_-$ contour so as to avoid the
singularities in the functions $\bs{S}$ on the right side of
Eq.~(\ref{eq:S4-start}). In the region $k_+ >0$, all of the
singularities are in the $k_-$ lower half-plane, and so the contour
integration vanishes. Therefore, we need only consider the region
$k_+<0$.  For the first term in the brackets, we do the following: when
$k_+ <\ell_+$, we close the $k_-$ contour in the lower half-plane and
pick up the pole at $k_{-} = -i\varepsilon$; when $\ell_+ < k_+ <0$, we
close the $k_-$ contour in the upper half-plane and pick up the pole at
$k_{-}=\frac{k_{\perp}^{2}}{k_{+}} + i\varepsilon$. For the second,
third, and fourth terms in brackets in Eq.~(\ref{eq:S4-start}), we
close the $k_-$ contour in the upper half-plane and pick up the pole at
$k_{-} = \frac{k_{\perp}^{2}}{k_{+}} +i\varepsilon$.  The results of the
contour integrations are
%---------------
\begin{eqnarray}
%---------------
\label{eq:S4-start-2}%
\bs{S}_{(B)}(\ell_{+},\ell_{-})& =& 
-2\alpha_s C_F
\left(\frac{\mu^2e^{\gamma_\textrm{E}}}{4\pi}\right)^\epsilon
\int_{-\infty}^{0} \frac{dk_+}{(-k_+)}
\int \frac{d^{D-2}k_\perp}{(2\pi)^{D-2}}
\frac{1}{\bs{k}_\perp^2}
\nonumber \\
&&
\times
\bigg[
\theta(\ell_+-k_+)\bm{S}(\ell_+-k_+,\ell_-)
+
\theta(k_+-\ell_+)\bm{S}(\ell_+-k_+,\ell_--\tfrac{\bs{k}_\perp^2}{k_+})
\nonumber \\
&&
\quad
-\bm{S}(\ell_+-k_+,\ell_-)
-\bm{S}(\ell_+,\ell_--\tfrac{\bs{k}_\perp^2}{k_+})
+\bm{S}(\ell_+,\ell_-)
\bigg].
%---------------
\end{eqnarray}
%---------------
It is apparent that the rapidity divergences that appear as $k_+\to 0$
and as $k_+\to -\infty$ with $\bs{k}_\perp^2$ fixed cancel in
Eq.~(\ref{eq:S4-start-2}). This is as expected, since the upper
semi-infinite Wilson lines cancel against the lower semi-infinite Wilson
lines to produce finite-length Wilson lines. We note that the IR
divergences that appear as $\bs{k}_\perp^2 \rightarrow 0$ also cancel,
reflecting the fact that the soft function is IR finite.

Making the change of variables $\bs{k}_{\perp}^{2}=xk_{+}\ell_{-}$ and
splitting the $k_{+}$ integration region $[-\infty,0]$  into
$[-\infty,\ell_+]$ and $[\ell_+,0]$, we obtain
%---------------
\begin{eqnarray}
\label{eq:S4}
%---------------
\bs{S}_{(B)}(\ell_{+},\ell_{-})
&=&
\frac{\alpha_s C_F}{2\pi}
\frac{\left(\mu^2e^{\gamma_\textrm{E}}\right)^\epsilon}{\Gamma(1-\epsilon)}
(\ell_-)^{-\epsilon}
\int_{-\infty}^{0} dx
\frac{1}{(-x)^{1+\epsilon}}
\nonumber \\
&&
\times
\bigg\{
\int_{\ell_+}^0 \frac{dk_+}{(-k_+)^{1+\epsilon}}
\Big[
\bm{S}(\ell_+-k_+,\ell_-)
-
\bm{S}(\ell_+-k_+,\ell_-(1-x))
\nonumber \\
&&
\quad\quad\quad\quad\quad\quad\quad\quad
+\bm{S}(\ell_+,\ell_-(1-x))
-\bm{S}(\ell_+,\ell_-)
\Big]
\nonumber \\
&&
\quad
+
\int_{-\infty}^{\ell_+} \frac{dk_+}{(-k_+)^{1+\epsilon}}
\Big[\bm{S}(\ell_+,\ell_-(1-x))
-\bm{S}(\ell_+,\ell_-)\Big]
\bigg\}.
%---------------
\end{eqnarray}
%---------------
In Eq.~(\ref{eq:S4}), there are no divergences as $k_+\to 0$ or as
$\bs{k}_\perp^2\to 0$.  All of the UV poles come either from the
  region $x\to -\infty$ and/or the region $k_+\to -\infty$. We test
        for one-loop UV divergences by replacing the all-orders soft
        functions with LO soft-quark propagators. Then, we see that the
        $x\to -\infty$ region gives a UV pole only if the argument of
        $\bs{S}$ is independent of $x$ and that the $k_+\to -\infty$
        region gives a UV pole only if the argument of $\bs{S}$ is
        independent of $k_+$.  It follows that the second and third
  terms in the integrand of the $k_+$ integration with the range
  $[\ell_+,0]$ do not contribute UV poles.  We extract the UV-pole
  contributions from the other terms. Then, for the remaining finite
  integration over $k_+$, we make the variable change $k_+
  =x\ell_{+}$. The result is
%---------------
\begin{eqnarray}
\label{eq:S4UV}
%---------------
&&
\bs{S}_{(B)}^\textrm{UV}(\ell_{+},\ell_{-})
\nonumber \\
&=&
\frac{\alpha_s C_F}{2\pi}
\frac{1}{\epsilon_\textrm{UV}}
\Bigg\{
-
\left[
\frac{1}{\epsilon_\textrm{UV}}
+\log\left(\frac{\mu^2}{-\omega-i\varepsilon}\right)
\right]
\bm{S}(\ell_+,\ell_-)
-
\int_{-1}^0 dx
\frac{\bm{S}(\ell_+,\ell_-(1-x))
-\bm{S}(\ell_+,\ell_-)}
{x}
\nonumber \\
&&
\quad\quad\quad\quad\quad
+
\int_{0}^{1} dx 
\frac{
\bm{S}(\ell_+(1-x),\ell_-)
-\bm{S}(\ell_+,\ell_-)}
{x}
-
\int_{-\infty}^{-1} dx
\frac{\bm{S}(\ell_+,\ell_-(1-x))}
{x}
\Bigg\}.
%---------------
\end{eqnarray}
%---------------
Here, we have inserted a term $-i\varepsilon$ into the argument of the
logarithm. This prescription is not necessary for the current case
$\omega=\ell_+\ell_-<0$. However, with this change, we can see that
Eq.~(\ref{eq:S4UV}) actually gives the correct analytic continuation of
the function on the right side for all $\omega$ in the complex
plane. Because the factors $(1-x)$ in the arguments of $\bs{S}$ are
non-negative over all of the ranges of integration, the right side of
Eq.~(\ref{eq:S4UV}) has the property that it is analytic for all
$\omega$, except for a cut along positive real axis that lies just below
the axis, which is the correct analyticity structure for the soft
function on the left side of Eq.~(\ref{eq:S4UV}).

In the application of the soft function to the process
$H\to \gamma\gamma$ through a $b$-quark loop, one
integrates the soft function over $\ell_\perp$ and takes the discontinuity.
The analytic continuation in Eq.~(\ref{eq:S4UV}) is not suited for this
purpose because, if one carries out the integration over $\ell_\perp$
and takes the discontinuity inside the infinite-range $x$ integration, the
resulting $x$ integration is divergent. This is easily seen 
from the LO expression for $\bs{S}$ in Eq.~(\ref{eq:SLO-exp}).

In order to remedy this situation, we rewrite the expression in
Eq.~(\ref{eq:S4UV}). First, we write the integrations over negative
values of $x$ as
%---------------
\begin{eqnarray}
%---------------
\label{eq:S4-rearrange}
&&-\int_{-\infty}^{-1}\frac{dx}{x}{\bs
  S}(\ell_{+},\ell_{-}(1-x))
-\int_{-1}^{0}\frac{dx}{x}~\left[\bs{S}(\ell_{+},\ell_{-}(1-x))-\bs{
    S}(\ell_{+},\ell_{-})\right]\nonumber\\
&=&\lim_{\delta\to 0}\left[
-\int_{-\infty}^{-\delta}\frac{dx}{x} \bs{S}(\ell_{+},\ell_{-}(1-x))
-\int_{-1}^{-\delta}\frac{dx}{x} \bs{S}(\ell_{+},\ell_{-})\right].
%---------------
\end{eqnarray}
%---------------
We consider the first term in brackets on the right side of
Eq.~(\ref{eq:S4-rearrange}). There are two cases: (i) when $0\leq
\mathrm{arg}(\omega)<\pi$, $\bs{S}(\ell_{+},\ell_{-}(1-x))$ has a cut
that extends from $x=1$ into the $x$ lower half-plane and is otherwise
analytic, (ii) when $-\pi< \mathrm{arg}(\omega)<0$,
$\bs{S}(\ell_{+},\ell_{-}(1-x))$ has a cut that extends from $x=1$ into
the $x$ upper half-plane and is otherwise analytic.  For case (i)
[(ii)], we deform the $x$ contour of integration into the upper [lower]
half-plane into a semicircle at infinity, a line from $\infty$ to
$\delta$, and a small semicircle from $\delta$ to $-\delta$ that is
traversed in the counterclockwise [clockwise] direction.  The
contribution of the semicircle at infinity vanishes, and the
contribution of the small semicircle is $\pm i\pi S(\omega)$, where the
upper [lower] sign corresponds to case (i) [(ii)]. For the second term
in brackets on the right side of Eq.~(\ref{eq:S4-rearrange}), we make
the change of variables $x\to -x$. With these transformations, the
expression in Eq.~(\ref{eq:S4UV}) becomes
%---------------
\begin{eqnarray}
\label{eq:S4UV2}
%---------------
\bs{S}_{(B)}^\textrm{UV}(\ell_{+},\ell_{-})
&=&
\frac{\alpha_s C_F}{2\pi}
\frac{1}{\epsilon_\textrm{UV}}
\Bigg\{
-
\left[
\frac{1}{\epsilon_\textrm{UV}}
+\log\left(\frac{\mu^2}{\omega+i\varepsilon}\right)
\right]
\bm{S}(\ell_+,\ell_-)
+
\int_{1}^\infty dx
\frac{\bm{S}(\ell_+,\ell_-(1-x))}
{x}
\nonumber \\
&&
\quad\quad\quad\quad\quad
+
\int_{0}^{1} dx 
\frac{
\bm{S}(\ell_+(1-x),\ell_-)
+\bm{S}(\ell_+,\ell_-(1-x))
-2\bm{S}(\ell_+,\ell_-)}
{x}
\Bigg\}.
\phantom{XX}
%---------------
\end{eqnarray}
%---------------
This expression is also a valid analytic continuation of
$\bs{S}_{(B)}^\textrm{UV}$ for all $\omega$, and it is suitable for use
in the application $H\to\gamma\gamma$ through a $b$-quark loop.\footnote{We
    note that one can obtain the expression in Eq.~(\ref{eq:S4UV2}) more
    directly for the case $\omega>0$ by carrying out the integration
    over $k_-$ in Eq.~(\ref{eq:S4-start}) with $\ell_->0$ and $\ell_+>0$.}
 
\subsection{Quark self-energy diagram}
There is also a contribution to the one-loop UV divergences that
arises from the one-loop quark self-energy diagram. As we have already
remarked, the UV divergence that is associated with the quark-mass
renormalization is removed by the standard QCD counterterm, and only the
wave-function-renormalization divergence contributes to the
soft-operator renormalization. It is given by
%---------------
\begin{eqnarray}
%---------------
\label{eq:S5UVxi}
\bs{S}_{(C)}^{\text{UV}}(\ell_{+},\ell_{-})& =& -\frac{\alpha_{s}C_{F}}{4\pi}\frac{1}{\epsilon_{\text{UV}}}\bs{S}(\ell_{+},\ell_{-}).
%---------------
\end{eqnarray}
%---------------
%\newpage
\subsection{$\textbf{\textit{Z}}_{\!\textbf{\textit{S}}}$ at one-loop
  order} 
Now let us summarize the one-loop contributions to the UV poles 
of the integrated soft function
$\bs{S}(\ell_{+},\ell_{-})$ for the case $\ell_{+}>0$, $\ell_{-}>0$:
%---------------
\begin{eqnarray}
\label{eq:total-UV}
%---------------
&&
\bm{S}^\textrm{UV}(\ell_+,\ell_-)
\nonumber \\
&=&
\bm{S}^\textrm{UV}_{(A_1)}(\ell_+,\ell_-)
+
\bm{S}^\textrm{UV}_{(A_3)}(\ell_+,\ell_-)
+
\bm{S}^\textrm{UV}_{(B)}(\ell_+,\ell_-)
+
\bm{S}^\textrm{UV}_{(C)}(\ell_+,\ell_-)
\nonumber \\
&=&
\frac{\alpha_s C_F}{2\pi}
\frac{1}{\epsilon_\textrm{UV}}
\int_{\bs{\ell}_{\perp}}
\Bigg[
\bm{S}(\ell_+,\ell_-,\ell_\perp)
\left(1
-\frac{n\sl_2n\sl_1}{4}
+\frac{n_2\cdot \ell}{\omega}\frac{n\sl_1\ell\sl_\perp}{2}\right)
\nonumber \\
&&
\quad\quad\quad\quad\quad\quad
+
\int_0^\infty
dx
\bigg\{
\frac{\bm{S}(\ell_+(1+x),\ell_-,\ell_\perp)}
{x(1+x)}
\left[1
-\frac{n\sl_2n\sl_1}{4}+
\frac{n_2\cdot \ell}{(1+x)\omega}\frac{n\sl_1\ell\sl_\perp}{2}
+\frac{m_b(n_2\cdot \ell)}{\omega}\frac{n\sl_1}{2}\right]
\nonumber \\
&&
\quad\quad\quad\quad\quad\quad\quad\quad\quad\quad\quad
-
\frac{\bm{S}(\ell_+,\ell_-,\ell_\perp)}
{x(1+x)}
\left[
1
-\frac{n\sl_2n\sl_1}{4}
+\frac{n_2\cdot \ell}{\omega}\frac{n\sl_1\ell\sl_\perp}{2}
+\frac{m_b(n_2\cdot \ell)}{\omega}\frac{n\sl_1}{2}
\right]
\bigg\}
\Bigg]\nonumber \\
&&
+
\frac{\alpha_s C_F}{2\pi}
\frac{1}{\epsilon_\textrm{UV}}
\int_{\bs{\ell}_{\perp}}
\Bigg[
\left(
1-\frac{n\sl_2 n\sl_1}{4}+\frac{n_1\cdot \ell}{\omega}
\frac{\ell\sl_\perp n\sl_2}{2}\right)
\bm{S}(\ell_{+},\ell_{-},\ell_\perp)
\nonumber \\
&&
\quad\quad\quad\quad\quad\quad
+
\int_0^\infty dx 
\bigg\{
\left[
1-\frac{n\sl_2 n\sl_1}{4}+\frac{n_1\cdot \ell}{(1+x)\omega}
\frac{\ell\sl_\perp n\sl_2}{2}
+\frac{m_b(n_1\cdot \ell)}{\omega}\frac{n\sl_2}{2}\right]
\frac{
\bm{S}(\ell_{+},(1+x)\ell_{-},\ell_\perp)}
{
x(1+x)}
\nonumber \\
&&
\quad\quad\quad\quad\quad\quad\quad\quad\quad\quad
-
\left[
1-\frac{n\sl_2 n\sl_1}{4}+\frac{n_1\cdot \ell}{\omega}\frac{\ell\sl_\perp n\sl_2}{2}
+\frac{m_b(n_1\cdot \ell)}{\omega}\frac{n\sl_2}{2}\right]
\frac{
\bm{S}(\ell_{+},\ell_{-},\ell_\perp)}
{x(1+x)}
\bigg\}
\Bigg]
\nonumber \\
&&
+
\frac{\alpha_s C_F}{2\pi}
\frac{1}{\epsilon_\textrm{UV}}
\Bigg\{
-
\left[
\frac{1}{\epsilon_\textrm{UV}}
+\log\left(\frac{\mu^2}{\omega}\right)
\right]
\bm{S}(\ell_+,\ell_-)
+
\int_{1}^\infty dx
\frac{\bm{S}(\ell_+,\ell_-(1-x))}
{x}
\nonumber \\
&&
\quad\quad\quad\quad\quad
+
\int_{0}^{1} dx 
\frac{
\bm{S}(\ell_+(1-x),\ell_-)
+\bm{S}(\ell_+,\ell_-(1-x))
-2\bm{S}(\ell_+,\ell_-)}
{x}
\Bigg\}
\nonumber \\
&&
-\frac{\alpha_s C_F}{4\pi}\frac{1}{\epsilon_\textrm{UV}}
\int_{\bm{\ell}_\perp}
\bm{S}(\ell_+,\ell_-,\ell_\perp).
%---------------
\end{eqnarray}
%---------------
We note that this expression contains contributions in which the
    argument of $\bs{S}$ is shifted away from
    $(\ell_+,\ell_-,\ell_\perp)$. As we have seen, these nonlocal
    contributions arise because of unorthodox momentum routing in the
    soft function that is required to factor the soft function from
        the jet functions.

Since, $\bs{S}(\ell_+,\ell_-)$ has no singularities on the real
$\omega=\ell_+\ell_-$ axis for $\omega<0$, the contribution in
Eq.~(\ref{eq:total-UV}) that is proportional to $\int_{1}^\infty dx
\bm{S}(\ell_+,\ell_-(1-x))/x$ gives a vanishing contribution to the
discontinuity of $\bs{S}^{\rm UV}(\ell_+,\ell_-)$. Therefore, we drop
this contribution in subsequent discussions.

We can express the discontinuity of the renormalized soft function
$S^{\text{R}}(\ell_{+},\ell_{-})$ in terms of the discontinuities of
its structure functions. We wish to renormalize the structure
    functions that are given in
    Eq.~(\ref{eq:integrated-soft-function-form-factor}). However, the
    $\ell_\perp$-dependent terms in Eq.~(\ref{eq:total-UV}) mix
    additional structure functions into the structure functions in
    Eq.~(\ref{eq:integrated-soft-function-form-factor}). These
    additional structure functions are
%---------------
\begin{eqnarray}
%---------------
\bm{S}_i(\omega)
=
\displaystyle
\int_{\bm{\ell}_\perp}\bm{S}_{i}(\omega,\bm{\ell}_\perp^2)
\frac{\bm{\ell}_\perp^2}{\omega},
\quad\textrm{for $i=5$, $6$, $7$, and $8$}.
%---------------
\end{eqnarray}
%---------------
We note that the renormalizations of these additional structure
    functions involve new structure functions, and so on, {\it ad
      infinitum}. We do not write out the renormalizations of
    these additional structure functions.

Using Eq.~(\ref{eq:total-UV}) as a starting point, we can write the
renormalized forms of the discontinuities of the structure functions
$S_1^R$--$S_4^R$ as convolution integrals.  We make the following
changes of integration variables: $\omega^\prime=(1+x)\omega$ for
$\bs{S}_{(A_1)}^{\rm UV}$ and $\bs{S}_{(A_3)}^{\rm UV}$, and
$\omega^\prime=(1-x)\omega$ for $\bs{S}_{(B)}^{\rm UV}$. The result is
%---------------
\begin{eqnarray}
%---------------
\label{eq:convolution}%
S_i^\textrm{R}(\omega)
=\sum_{j=1}^8 
\int_0^\infty d\omega'
Z_S^{ij}(\omega,\omega';\mu)
S_j(\omega'),
\quad \textrm{for $i=$1, 2, 3, and 4,}
%---------------
\end{eqnarray}
%---------------
where
%---------------
\begin{eqnarray}
%---------------
Z_S^{ij}
(\omega,\omega';\mu)
=\delta(\omega-\omega')\delta^{ij}
+
\frac{\alpha_sC_F}{4\pi}
\frac{1}{\epsilon_\textrm{UV}}
\bm{M}_S^{ij}(\omega,\omega';\mu),
%---------------
\end{eqnarray}
%---------------
and the matrix representation of $\bm{M}_S^{ij}(\omega,\omega';\mu)$
is given by
%---------------
\begin{eqnarray}
%---------------
\label{eq:renorm-matrix}%
&&
\bm{M}_S(\omega,\omega';\mu)
\nonumber \\
&&
\!\!\!\!\!\!\!\!\!
=
\begin{pmatrix}
d-2a-2b-2c & 0 & 0 & 0 & 0 & 0 & 0 & 0 
\\
-\frac{m_b^2}{\omega}b 
& d-a-\frac{\omega'}{\omega}b-\frac{\omega'+\omega}{\omega}c
&0 &0 & -(a+b) & 0 &0 & 0
\\
-\frac{m_b^2}{\omega}b & 0
&d-a-\frac{\omega'}{\omega}b-\frac{\omega'+\omega}{\omega}c 
&0 & -(a+b)& 0 & 0 & 0
\\
2(a+b) & -\frac{\omega'}{\omega}b 
& -\frac{\omega'}{\omega}b & d-2c &
0 &a+b & a+b & 0
\end{pmatrix},
\nonumber \\
%---------------
\end{eqnarray}
%---------------
with $a$, $b$, $c$, and $d$ defined by
%---------------
\begin{eqnarray}
%---------------
a &=& 2\delta(\omega-\omega'),
\nonumber \\
b &=& 2\left[\frac{\omega\theta(\omega'-\omega)}{\omega'(\omega'-\omega)}
\right]_+,
\nonumber \\
c&=&
2
\left[
\frac{\theta(\omega-\omega')}{\omega-\omega'}
\right]_+,
\nonumber \\
d &=&
\left[
\frac{2}{\epsilon_\textrm{UV}}
+2\log\left(\frac{\mu^2}{\omega}\right)+1
\right]\delta(\omega-\omega').
%---------------
\end{eqnarray}
%---------------
Here, the plus distribution is defined by
%---------------
\begin{eqnarray}
%---------------
\int_{0}^{\infty}d\omega^{\prime}\frac{f(\omega^{\prime})}{[g(\omega^{\prime})]_{+}} = \int_{0}^{\infty}d\omega^{\prime}\frac{f(\omega^{\prime}) - f(\omega)}{g(\omega^{\prime})}.
%---------------
\end{eqnarray}
%---------------
The plus distributions correspond to the nonlocal contributions in
    Eq.~(\ref{eq:total-UV}), while the $\delta$ function corresponds to
    the local contributions.

In Eq.~(\ref{eq:convolution}), we have written down the
      renormalization of the soft function for a general Dirac
      structure. Only the structure function $S_1$ is relevant for the
      process $H\rightarrow \gamma\gamma$ through a $b$-quark loop
      because the radiative jet functions in $\langle \gamma\gamma|{\cal
        O}_3|H \rangle$ produce a factor $\slashed{n}_2$ on the left of
      the soft function and a factor $\slashed{n}_1$ on the right of the
      soft function, which project out the structure function
      ${S}_1$. However, we have included the other Dirac structures in
      Eq.~(\ref{eq:convolution}) because of the possibility that they
      might be relevant for exclusive processes other than $H\rightarrow
      \gamma\gamma$ through a $b$-quark loop.

Because the matrix in Eq.~(\ref{eq:renorm-matrix}) mixes
${S}_5$--${S}_8$ with ${S}_1$--${S}_4$, it is not possible to write a
closed-form evolution equation for all of ${S}_1$--${S}_4$.  Therefore,
we specialize to the case of ${S}_1$, for which there is no mixing in
the renormalization. For ${S}_1$, we have
%---------------
\begin{eqnarray}
%---------------
Z_{S}^{11}
&=& 
\delta(\omega^{\prime} -\omega) +\frac{\alpha_{s}C_{F}}{4\pi}
\frac{1}{\epsilon_{\text{UV}}}\left(d-2a-2b-2c\right)
\nonumber \\
&=&  
\delta(\omega^{\prime} -\omega) + \frac{\alpha_{s}C_{F}}{4\pi}
\frac{1}{\epsilon_{\text{UV}}}
\bigg\{
\left[\frac{2}{\epsilon_{\text{UV}}} 
+ 2 \log\left(\frac{\mu^{2}}{\omega}\right)-3\right]\delta(\omega^{\prime}-\omega)
\nn\\
&& 
\quad\quad\quad\quad\quad\quad\quad\quad\quad\quad
- 4 \omega\left[\frac{\theta(\omega^{\prime}-\omega)}{\omega^{\prime}(\omega^{\prime}-\omega)}+\frac{\theta(\omega-\omega^{\prime})}{\omega(\omega-\omega^{\prime})}\right]_{+}
\bigg\}.
%---------------
\end{eqnarray}
%---------------
This result confirms the conjecture in Ref.~\cite{Liu:2020eqe} for the
renormalization of the soft function in order $\alpha_s$. It is
consistent with the explicit calculation of the soft function in order
$\alpha_s$ that is given in Eqs.~(4.7)--(4.8) of
Ref.~\cite{Liu:2019oav}.  Note, however, that one cannot deduce the form
of the one-loop renormalization and evolution of the soft function from
the order-$\alpha_{s}$ contribution to the soft function because,
      once the integration over $\omega^\prime$ in
      Eq.~(\ref{eq:convolution}), has been carried out, the nonlocal
      renormalization factor $Z_S$ cannot be reconstructed.
 
\subsection{One-loop evolution equation}
We obtain the evolution equation for $S_1$ by differentiating
    Eq.~(\ref{eq:convolution}) with respect to the renormalization scale
    $\mu$:
\begin{eqnarray}
\frac{d}{d\log{\mu}}S_{1}^\textrm{R}(\omega,\mu) 
= -\int_{0}^{\infty}d\omega^{\prime}\,
\gamma_{S}(\omega,\omega^{\prime},\mu)S_{1}^\textrm{R}(\omega^{\prime},\mu),
\end{eqnarray}
where
\begin{eqnarray}
\gamma_{S}(\omega,\omega^{\prime},\mu) &=&-\frac{d\log{Z_{S}^{11}}}{d\log{\mu}} \nn\\
&=& -\frac{\alpha_{s}C_{F}}{4\pi}\Bigg\{\left[4\log{\left(\frac{\omega}{\mu^{2}}\right)}+6\right]\delta\left(\omega^{\prime}-\omega\right)\nn\\
&&
\quad\quad\quad\quad
+8\omega\left[\frac{\theta(\omega^{\prime}-\omega)}{\omega^{\prime}(\omega^{\prime}-\omega)}+\frac{\theta(\omega-\omega^{\prime})}{\omega(\omega-\omega^{\prime})}\right]_{+}\Bigg\}+\mathcal{O}\left(\alpha_{s}^{2}\right),
\end{eqnarray}
and we have made use of the order-$\alpha_{s}$ dependence of
    $\alpha_s$ on $\mu$, which is given by
\begin{eqnarray}
\frac{d\alpha_{s}}{d\log{\mu}} %= -2\epsilon\alpha_{s}+2\beta\alpha_{s} 
= -2\epsilon\alpha_{s}+\mathcal{O}\left(\alpha_{s}^{2}\right).
\end{eqnarray}
This result confirms the conjecture in Ref.~\cite{Liu:2020eqe} for the
    one-loop evolution of the soft function.

\section{Summary \label{sec:summary}}

In this paper we have presented a method for computing the
renormalization and evolution of the soft function that
appears in the factorization theorem for Higgs-boson decay to two
photons through a $b$-quark loop \cite{Liu:2019oav}. The renormalization
is not straightforward because the factorization of the soft function
from the jet functions leads to an unorthodox routing of loop momenta in
the soft function. That unorthodox momentum routing results in
nonlocal contributions to the renormalization and evolution of the soft
function.

Our analysis of the renormalization of the soft function makes use of
its analyticity properties. In Ref.~\cite{Liu:2019oav}, it was asserted
that the soft function is analytic in the $\omega$ complex plane, except
for singularities on the nonnegative real axis. ($\omega$ is the
    product of the longitudinal momenta that are carried by the soft
    function.) Somewhat surprisingly, the singularities extend to
values of $\omega$ that are less than $m_b^2$. We have shown, through
explicit examples in light-front perturbation theory, that such
singularities arise because a $b$-quark line can carry an infinite
longitudinal loop momentum, resulting in the vanishing of the $b$-quark
intermediate-state light-front energy.  We have also argued, making use
of light-front perturbation theory, that the soft function is analytic
everywhere in the complex $\omega$ plane, except for the nonnegative
real axis, thus confirming the assertion in Ref.~\cite{Liu:2019oav}.

We have given an explicit calculation of the one-loop renormalization
and evolution of the soft function. Our result for the one-loop
renormalization is consistent with the calculation of the
order-$\alpha_{s}$ contribution to the soft function in
Ref.~\cite{Liu:2019oav}. Note, however, that one cannot deduce the
      nonlocal form of the one-loop renormalization/evolution of the
      soft function from the order-$\alpha_{s}$ contribution to the soft
      function because the renormalization/evolution involves a
      convolution with respect to $\omega$ of the nonlocal
      renormalization factor $Z_S$ with the soft function. One must make
      the convolution explicit in order to deduce $Z_S$.

Our results for the one-loop renormalization and evolution of the soft
    function confirm the conjectured one-loop form in
    Ref.~\cite{Liu:2020eqe}. This puts on a solid footing the one-loop
    contribution to the evolution kernel for the soft function, which is
    used to resum logarithms of $m_{H}^2/m_b^2$.  The two-loop
    contribution to evolution kernel would be required to achieve
    greater precision in the resummation of logarithms of
    $m_{H}^2/m_b^2$.  The form of the two-loop renormalization
    renormalization of the soft function has also been conjectured in
    Ref.~\cite{Liu:2020eqe}, and it is important to check that
    conjecture through explicit calculations.  In principle, the methods
    that we have presented generalize straightforwardly to higher-order
    calculations of the renormalization and evolution of the soft
    function. However, because it is necessary to single out the
    longitudinal components of the loop momenta in order to capture the
    nonlocal nature of the renormalizations, one cannot use standard
    two-loop methods in such calculations, and the calculations may be
    technically challenging.

The unorthodox momentum routing in the soft function that we have
    noted in this paper arises because the soft function contains a
    soft-quark line, in contrast with the purely gluonic soft functions
    that arise in typical leading-power factorization theorems. The
    soft-quark line appears because the physical process first occurs at
    subleading power in the quark mass ($m_b^1$). At subleading
    power in the quark mass, a soft-quark pinch singularity can
    give a relative order-one contribution \cite{Bodwin:2014dqa}. Hence,
    we expect this phenomenon to appear in any exclusive process that
    proceeds through a quark helicity flip, and the methods that we have
    presented in this paper should be applicable in those
    situations. For example, those methods may be relevant to the
    renormalization of soft function in the, as yet unproven,
    factorization theorem for the amplitude for $e^+e^-\to J/\psi
    +\eta_c$.

\begin{acknowledgments}

We wish to acknowledge the contributions of Hee Sok Chung to work on
related topics at an early stage in this project.  We thank Ze Long Liu
and Jian Wang for several useful discussions.  The work of G.T.B.\ and
X.-P.W.\ is supported by the U.S.\ Department of Energy, Division of
High Energy Physics, under Contract No.\ DE-AC02-06CH11357.  The work of
J.L. is supported by the National Research Foundation of
Korea under Contract No. NRF-2020R1A2C3009918. 
The work of J.-H.E. is supported by the National Natural Science 
Foundation of China (NSFC) through Grant No. 11875112.
 The submitted manuscript
has been created in part by UChicago Argonne, LLC, Operator of Argonne
National Laboratory. Argonne, a U.S.\ Department of Energy Office of
Science laboratory, is operated under Contract
No.\ DE-AC02-06CH11357. The U.S.\ Government retains for itself, and
others acting on its behalf, a paid-up nonexclusive, irrevocable
worldwide license in said article to reproduce, prepare derivative
works, distribute copies to the public, and perform publicly and display
publicly, by or on behalf of the Government.

All authors contributed equally to this work.

\end{acknowledgments}

\appendix

\section{Analyticity of the soft function: examples\label{sec:analyticity2}}

In this appendix, we present some one-loop examples to illustrate the
analyticity properties of the soft function. We consider all of the
one-loop contributions to the soft function.  We find that some of the
diagrams contain denominators that can vanish for all $\omega\geq 0$,
and, hence, contribute to an imaginary part. We also find, as expected
from our general argument in Sec.~\ref{sec:analyticity}, that none of
the denominators can vanish if $\omega$ is negative.

Our arguments rely on the use of light-front perturbation theory, whose
Feynman rules we now summarize.  In light-front perturbation theory,
expressions for Feynman diagrams consist of (i) vertices, whose Feynman
rules are the same as in covariant perturbation theory, (ii) energy
denominators for each intermediate state, which consist of the incoming
light-front energy minus the on-shell light-front energies of the
particles that are included in the intermediate state, (iii) factors of
the inverse of the longitudinal ($+$) momenta of the intermediate-state
particles, and (iv) integrations over the $+$ and transverse components
of the loop momenta \cite{Collins:2011zzd, Chang:1968bh, Kogut:1969xa,
  Yan:1973qg}.  The vertices are assigned an ordering in $x_+$. For each
intermediate-state line, the allowed propagation is from lower to higher
$x_+$, and, so the longitudinal component of momentum is positive. We
have drawn the figures in this appendix to indicate the light-front time
orderings. We note that the points $X$ and $Z$ are at equal light-front
time.

In our case, there are also Wilson lines present. The $n_1$ Wilson line
is instantaneous in light-front time. That is, it does not
propagate. Therefore, it simply produces an overall factor. In the case
of the $n_2$ Wilson line, the open end is always at light-front time
$-\infty$ and the vertex end is at some finite light-front time. It has
intermediate-state energy zero.

The analyticity of the soft function is described by the analyticity of
its structure functions. As we have seen, the structure functions are
functions of the product $\ell_+\ell_-$.  Therefore, in considering the
analyticity of the structure functions as a function of
$\omega=\ell_+\ell_-$, we can, without loss of generality, take $\ell_+
>0$.

%--------------------------------------
\begin{figure}
\centering                          
\includegraphics[width=0.25\columnwidth]{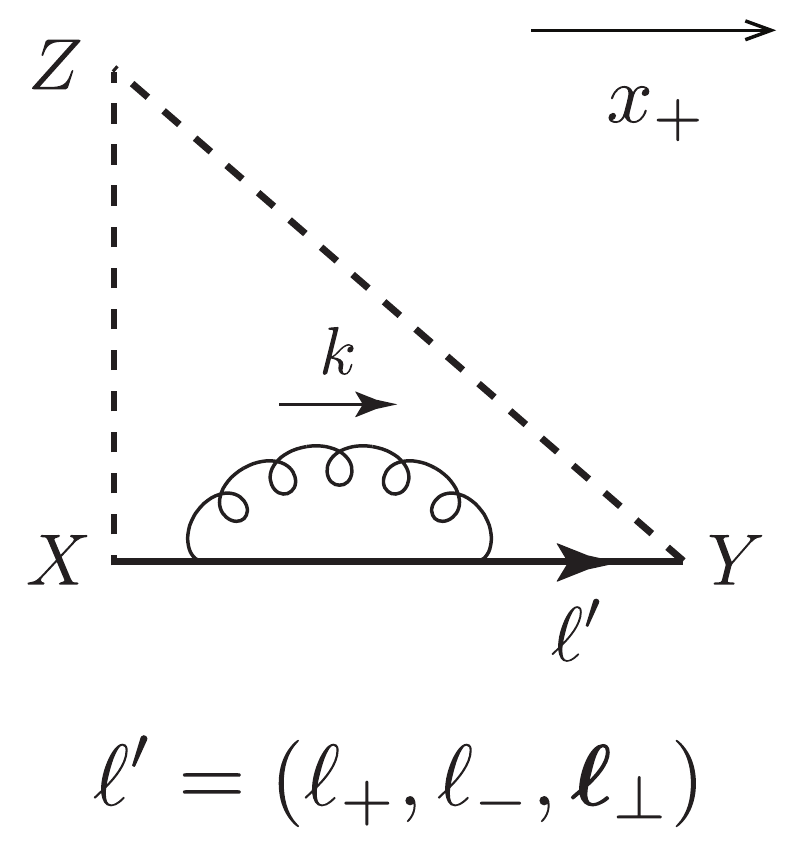}
\caption{\label{fig:self-energy} Quark self-energy diagram.}
\end{figure}
%--------------------------------------
Let us begin by considering the quark self-energy diagram in
    Fig.~\ref{fig:self-energy}. The corresponding propagator
    denominators in covariant perturbation theory are given by
\begin{eqnarray}
&&\frac{1}{k_+k_--\bs{k}_\perp^2+i\ve} 
\frac{1}{(\ell_+-k_+)(\ell_--k_-)-(\bs{\ell}_\perp-\bs{k}_\perp)^2-m_b^2+i\ve}.
\end{eqnarray}
The $k_-$ contour integration gives a nonzero result only if
$0<k_+<\ell_+$. We close the $k_-$ contour of integration in the lower
half-plane, picking up the gluon pole at $k_-=\bs{k}_\perp^2/k_+-i\ve$, to
obtain a result that is proportional to
\begin{eqnarray}
-2\pi i \theta(k_+)\theta(\ell_+-k_+) \frac{1}{k_+} \frac{1}{\ell_+-k_+}
\frac{1}{\ell_--\bs{k}_\perp^2/k_+-
[(\bs{\ell}_\perp-\bs{k}_\perp)^2+m_b^2]/(\ell_+-k_+) +i\ve}.
\end{eqnarray}
This expression has a simple interpretation in light-front perturbation
theory. The $\theta$ functions enforce the positivity of the $+$
light-front longitudinal momenta. The next two factors are the
light-front normalizations of the gluon and soft-quark propagators,
respectively. The denominator in the last factor is the light-front
energy denominator from the intermediate state that consists of the
gluon and the soft-quark. In the energy denominator, $\ell_-$ is the
light-front energy, $\bs{k}_\perp^2/k_+$ is the gluon intermediate-state
energy, and $[(\bs{\ell}_\perp-\bs{k}_\perp)^2+m_b^2]/(\ell_+-k_+)$ is
the soft-quark intermediate-state energy. We see that, owing to the
constraints that are imposed by the positivity of the $+$ light-front
longitudinal momenta, $k_+$ is always finite. Consequently, the energy
denominator cannot vanish, and thereby produce an imaginary part, unless
$\omega=\ell_+\ell_-\geq m_b^2$.

%---------------------------------------
\begin{figure}
\centering                          
\includegraphics[width=0.55\columnwidth]{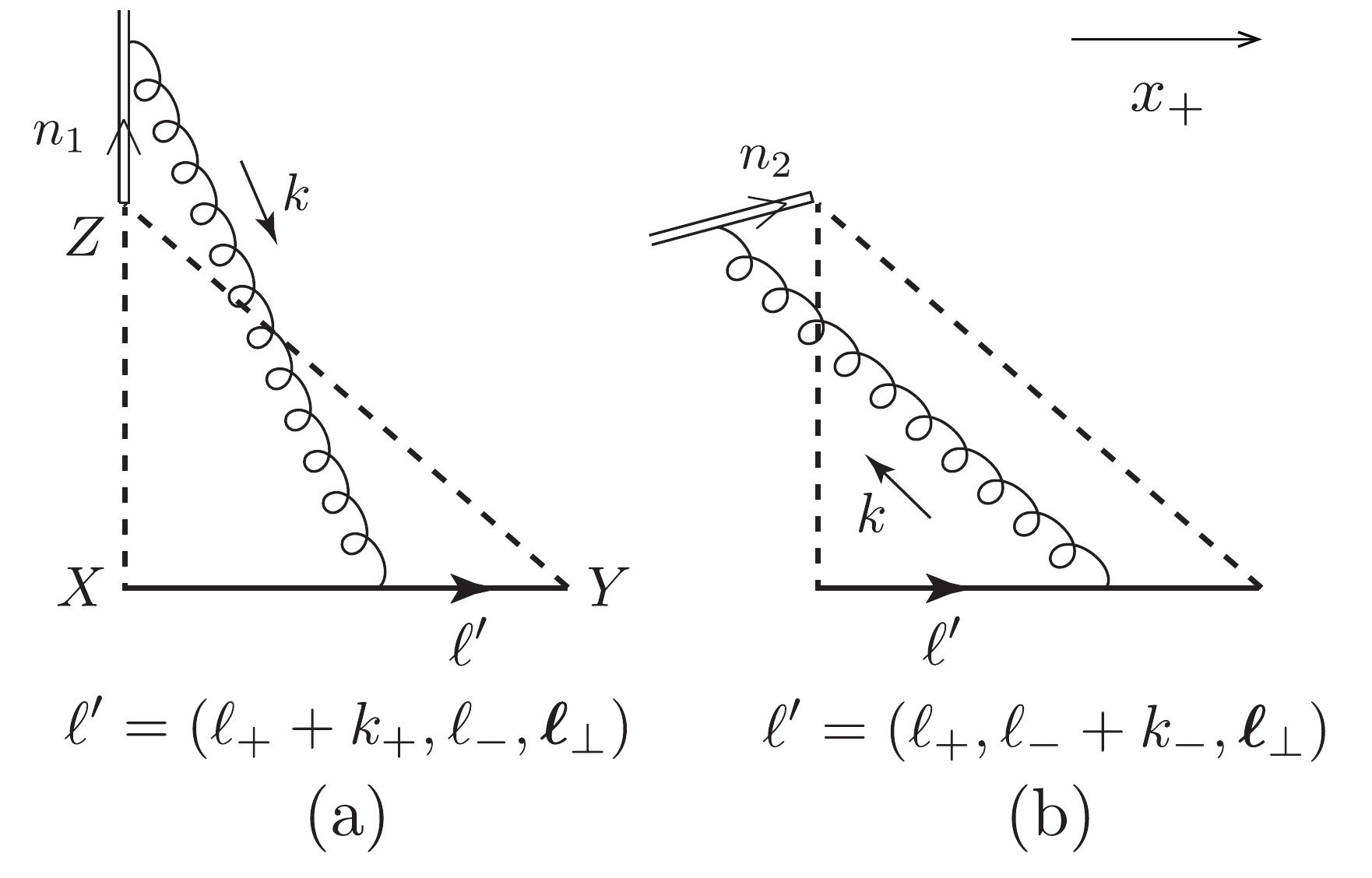}
\caption{\label{fig:analyticity1}Diagrams that illustrate the
  analyticity properties of the soft function, as is explained in the
  text.}
\end{figure}
%---------------------------------------

Next, we consider the diagram of Fig.~\ref{fig:analyticity1}(a).  The
corresponding propagator denominators in covariant perturbation theory
are given by
%---------------
\begin{eqnarray}
%---------------
\frac{1}{-k_{+} +i\ve} \frac{1}{k^2+i\ve}
\frac{1}{(\ell_+ +k_+)\ell_--\bs{\ell}_\perp^2-m_b^2+i\ve} 
\frac{1}{\ell_+(\ell_--k_-)-(\bs{\ell}_\perp-\bs{k}_\perp)^2-m_b^2+i\ve}.
\phantom{X}
%---------------
\end{eqnarray}
%---------------
For $\ell_{+}>0$, all of the poles in the $k_-$ complex plane are in the
upper half-plane unless $k_+ >0$. Hence, we close the $k_-$ contour of
integration in the lower half-plane, picking up the residue of the pole
from the gluon propagator at $k_-=\bs{k}_\perp^2/k_+-i\ve$ to obtain a
result that is proportional to
\begin{eqnarray}
\label{eq:analyticity1p}%
&&-2\pi i\theta(k_+)
\frac{1}{\ell_+} \frac{1}{\ell_++k_+} \frac{1}{k_+} 
\frac{1}{-k_{+}+i\ve}
\frac{1}{\ell_--\bs{k}_\perp^2/k_+
-[(\bs{\ell}_\perp-\bs{k}_\perp)^2+m_b^2]/\ell_++i\ve}\nn\\
&&\times \frac{1}{\ell_--(\bs{\ell}_\perp^2+m_b^2)/(\ell_+ +k_+)+i\ve}.
\end{eqnarray}
This expression also has a simple interpretation in light-front
perturbation theory: the $\theta$ function enforces the forward
propagation of the gluon; the next three factors are the light-front
normalizations of the left-hand and right-hand soft-quark propagators
and the gluon, respectively; the next factor comes from the
instantaneous Wilson-line propagator; the next factor is the energy
denominator from the intermediate state that consists of the gluon and
the left-hand soft-quark propagator; the last factor is the light-front
energy denominator from the intermediate state that consists of the
right-hand soft-quark line.  We see that in this case, in contrast with
the case of the quark self-energy diagram, the momentum $k_+$ can become
infinite without resulting in a negative longitudinal momentum in any of
the particle lines. This is a consequence of the unorthodox routing of
the momentum $k_+$, which appears in the right-hand soft-quark
propagator, rather than in the left-hand soft-quark propagator. Then,
for sufficiently large $k_+$, the second energy denominator can vanish
for $\ell_-\geq 0$, producing and imaginary part.  Note that the energy
denominator can never vanish for $\ell_-<0$ because all of the
intermediate-state energies are positive.

In the remaining examples in this Appendix, we couch the discussion in
    terms of light-front perturbation theory. However, it should be
    remembered that one can obtain the light-front expressions that
    appear in the discussions by integrating the
    covariant-perturbation-theory expressions over $k_-$, as we have
    done in the examples above.

A light-front analysis of the denominators of the diagram of
Fig.\ref{fig:analyticity1}(b) yields 
\begin{eqnarray}
&&-2\pi i \theta(-k_+) \frac{1}{\ell_+} 
\frac{1}{k_+}
\frac{1}{\ell_+-k_+} 
\frac{1}{\bs{k}_\perp^2/k_+}
\frac{1}{\ell_-+\bs{k}_\perp^2/k_+-(\bs{\ell}_\perp^2+m_b^2)/\ell_++i\ve}\nn\\
&&\times \frac{1}{\ell_--[(\bs{\ell}_\perp-\bs{k}_\perp)^2+m_b^2]/(\ell_+-k_+)
  +i\ve}.
\end{eqnarray}
Here, the denominator $\bs{k}_\perp^2/k_+$ is the energy denominator
from the left-most intermediate state, which consists of the gluon and
Wilson line. Because this intermediate state appears at an earlier
light-front time than the injection of the light-front energy $\ell_-$,
it has vanishing initial light-front energy. We note also that the
on-shell Wilson line in the intermediate state has light-front energy
zero. Because of the unorthodox momentum routing, $-k_+$ can reach
    positive infinity without causing of any of the particle lines to move
    backward. Consequently, for $-k_+$ sufficiently large, the
    intermediate state in the third energy denominator can have zero
    light-front energy, and it produces an imaginary part for
    $\ell_-\geq 0$. Since all of the intermediate-state light-front
    energies are positive, neither energy denominator can vanish if
$\omega=\ell_+\ell_-$ is negative.

\begin{figure}
\centering                          
\includegraphics[width=1\columnwidth]{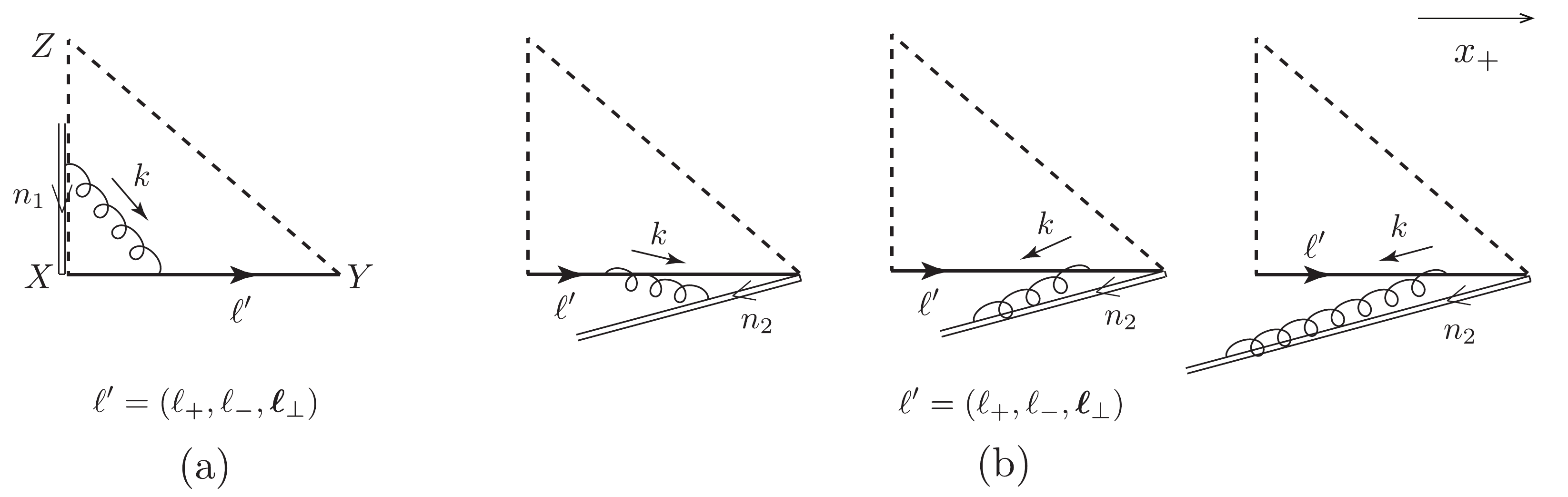}
\caption{\label{fig:analyticity1b}Diagrams that illustrate the
  analyticity properties of the soft function, as is explained in the text.}
\end{figure}

A light-front analysis of the denominators of the diagram of
Fig.~\ref{fig:analyticity1b}(a) yields 
\begin{eqnarray}
&&-2\pi i \theta(k_+)\theta(\ell_+-k_+) \frac{1}{\ell_+}
\frac{1}{k_+} \frac{1}{\ell_+-k_+} 
\frac{1}{-k_+ +i\ve}
\frac{1}{\ell_--\bs{k}_\perp^2/k_+
-[(\bs{\ell}_\perp-\bs{k}_\perp)^2+m_b^2]/(\ell_+-k_+)+i\ve} \nn\\
&&\times \frac{1}{\ell_--(\bs{\ell}_\perp^2+m_b^2)/\ell_++i\ve}.
\end{eqnarray}
The $\theta$-function constraints enforce the requirements that the
gluon, the left-hand, and the right-hand soft-quark lines be forward
moving. Because of these constraints, the longitudinal momentum $k_+$
can never become infinite in magnitude and, hence, the heavy quark can
never have an intermediate state with zero light-front
energy. Therefore, neither energy denominator can vanish for either sign
of $\omega=\ell_+\ell_-$.

%---------------------------------------
%---------------------------------------

A light-front analysis of the denominators of the diagrams of
Fig.~\ref{fig:analyticity1b}(b) yields 
\begin{eqnarray}
\label{eq:analyticity1b}%
&&-2\pi i\theta(k_+)\theta(\ell_+-k_+) \frac{1}{\ell_+} \frac{1}{k_+} 
\frac{1}{\ell_+-k_+}
\frac{1}{\ell_--(\bs{\ell}_\perp^2+m_b^2)/\ell_+ +i\ve}
\nonumber \\
&&\quad
\times \frac{1}{\ell_--\bs{k}_\perp^2/k_+-
[(\bs{\ell}_\perp -\bs{k}_\perp)^2+m_b^2]/(\ell_+-k_+)+i\ve}
\frac{1}{\ell_--[(\bs{\ell}_\perp -\bs{k}_\perp)^2+m_b^2]/
(\ell_+-k_+)+i\ve}
\nonumber \\
&&-2\pi i \theta(-k_+) \frac{1}{\ell_+} \frac{1}{k_+} 
\frac{1}{\ell_+-k_+}
\frac{1}{-\bs{k}_\perp^2/k_+} 
\frac{1}{\ell_--(\bs{\ell}_\perp^2+m_b^2)/\ell_+ +i\ve}
\nonumber \\
&&
\quad
\times \frac{1}{\ell_--[(\bs{\ell}_\perp -\bs{k}_\perp)^2+m_b^2]/
(\ell_+-k_+)+i\ve}.
\end{eqnarray}
In this case, there are three different contributions: the first figure
shows the ordering for $k_+>0$, and the last two figures show the two
orderings for $k_+<0$.\footnote{The two orderings for $k_+<0$ are
trivial because $k_-$ does not flow through the left-hand soft-quark
line. Consequently, sum of the contributions from these two orderings is
identical to the contribution that one would obtain by computing
independent energy denominators for the left-hand soft-quark line and
the left-hand gluon and Wilson lines.}  In the contribution in
Eq.~(\ref{eq:analyticity1b}) for which $k_+$ is positive, $k_+$ cannot
become infinite without causing the second soft-quark line to move
    backward. (This leads to the constraint $\ell_+-k_+>0$.) Hence,
the soft-quark lines can never have zero intermediate-state energy and
the energy denominators can never vanish.  In the contribution in
Eq.~(\ref{eq:analyticity1b}) for which $k_+$ is negative, $-k_+$ can
go to positive infinity without causing any of the particle lines to
    move backward. This is a consequence of the fact that the $n_2$
    Wilson line can carry the momentum component $k_+$, but is
    insensitive to it. It follows that the third energy denominator
can vanish when $\omega=\ell_+\ell_-\geq 0$ and, thereby, produce an
    imaginary part. As is usual, none of the light-front energy
denominators can vanish when $\omega=\ell_+\ell_-$ is negative.

%---------------------------------------
\begin{figure}
\centering                          
\includegraphics[width=0.90\columnwidth]{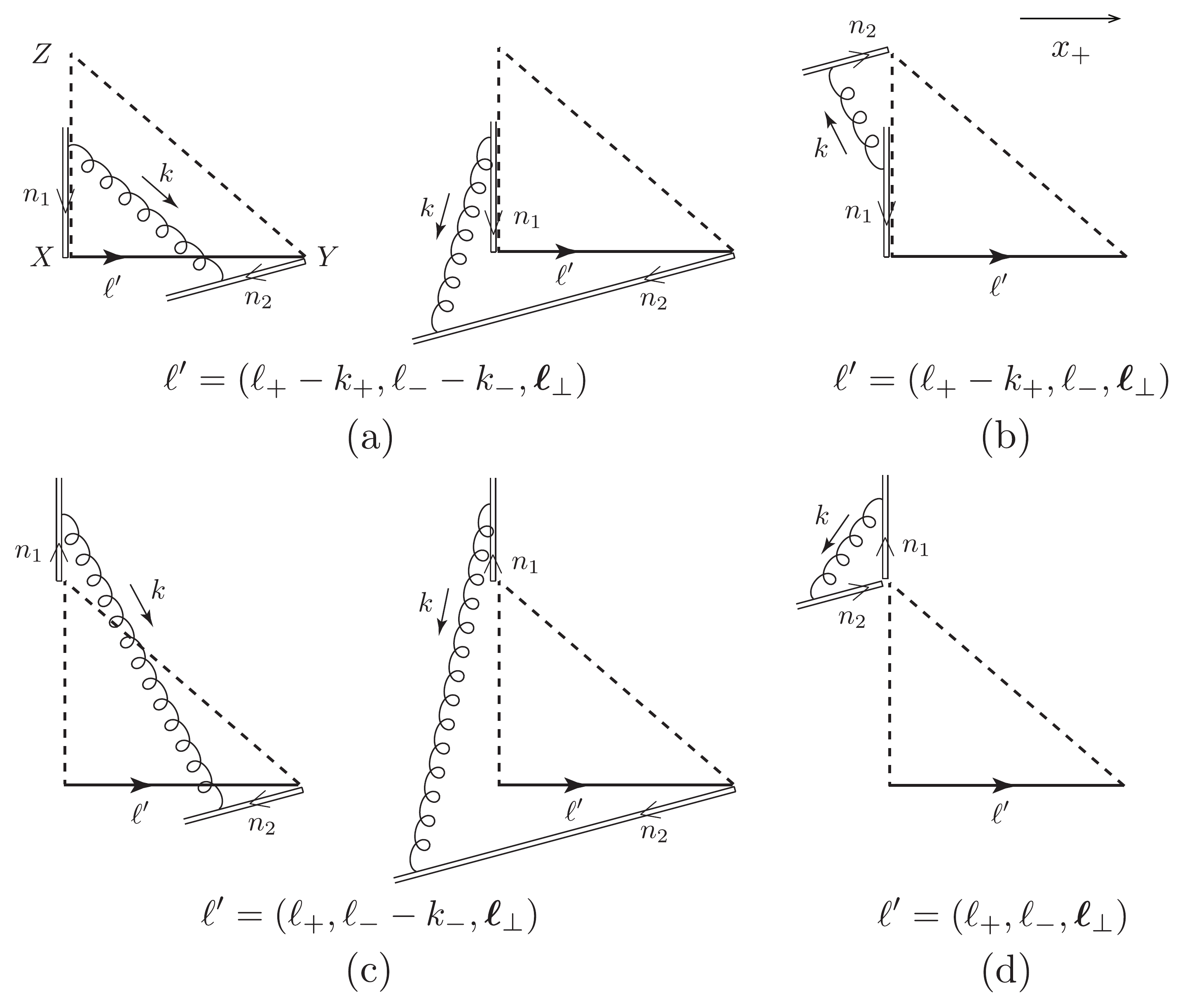}
\caption{\label{fig:analyticity2}Additional diagrams that illustrate
      the analyticity properties of the soft function, as is explained
      in the text.}
\end{figure}
%---------------------------------------

A light-front analysis of the denominators of the diagrams
Fig.~\ref{fig:analyticity2}(a) yields
\begin{eqnarray}
&&-2\pi i \theta(k_+)\theta(\ell_+-k_+)\frac{1}{k_+}
\frac{1}{\ell_+-k_+} \frac{1}{-k_+ +i\ve}
\frac{1}{\ell_--\bs{k}_\perp^2/k_+
-(\bs{\ell}_\perp^2+m_b^2)/(\ell_+-k_+)+i\ve} 
\nonumber \\
&&
\quad
\times  \frac{1}{\ell_--(\bs{\ell}_\perp^2+m_b^2)/(\ell_+-k_+) +i\ve}\nn\\
&&-2\pi i \theta(-k_+) \frac{1}{k_+} \frac{1}{\ell_+-k_+}
\frac{1}{-k_++i\ve} 
\frac{1}{-\bs{k}_\perp^2/k_+}
\frac{1}{\ell_--(\bs{\ell}_\perp^2+m_b^2)/(\ell_+-k_+)+i\ve}.
\end{eqnarray}
Again, there are two different contributions: the left-hand figure shows
the ordering for $k_+>0$, and the right-hand figure shows the ordering
for $k_+<0$.  In the contribution for which $k_+$ is positive, $k_+$
cannot become infinite without causing the soft-quark line to move
    backward. (This leads to the constraint $\ell_+-k_+>0$.) Hence,
the soft-quark line cannot have zero intermediate-state energy, and the
energy denominators can never vanish. However, for the contribution in
which $k_+$ is negative, $-k_+$ can become infinite. Again, this is a
    consequence of the fact that the $n_2$ Wilson line can carry the
    momentum component $k_+$, but is insensitive to it. It follows that
    the second energy denominator can vanish for $\omega=\ell_+\ell_->0$
    and produce an imaginary part.  Again, we see that none of the
light-front energy denominators can vanish when $\omega=\ell_+\ell_-$ is
negative.

A light-front analysis of the denominators of the diagram of
Fig.~\ref{fig:analyticity2}(b) yields
\begin{eqnarray}
&&2\pi i \theta(-k_+) \frac{1}{k_+} \frac{1}{\ell_+-k_+}
\frac{1}{-k_++i\ve} \frac{1}{\bs{k}_\perp^2/k_+} 
\frac{1}{\ell_--(\bs{\ell}_\perp^2+m_b^2)/(\ell_+-k_+)+i\ve}.
\end{eqnarray}
Considering the overall signs from the propagator and vertex factors
from $S_{n_2}$ and $S^\dagger_{n_2}$, we find that there is a
cancellation of the result from the diagram of
Fig.~\ref{fig:analyticity2}(b) with the second term in the result from
the diagram of Fig.~\ref{fig:analyticity2}(a).

A light-front analysis of the denominators of the diagrams of
Fig.~\ref{fig:analyticity2}(c) yields
\begin{eqnarray}
&&-2\pi i\theta(k_+)\frac{1}{\ell_+} \frac{1}{k_+} \frac{1}{-k_+ +i\ve} 
\frac{1}{\ell_--(\bs{\ell}_\perp^2+m_b^2)/\ell_+-\bs{k}_\perp^2/k_+
  +i\ve}
\frac{1}{\ell_--(\bs{\ell}_\perp^2+m_b^2)/\ell_+ +i\ve}\nn\\
&&-2\pi i\theta(-k_+)\frac{1}{\ell_+} \frac{1}{k_+} \frac{1}{-k_+ +i\ve} 
\frac{1}{-\bs{k}_\perp^2/k_+}
\frac{1}{\ell_--(\bs{\ell}_\perp^2+m_b^2)/\ell_+ +i\ve}.
\end{eqnarray}
There are two different contributions: the left-hand figure shows the
ordering for $k_+>0$, and the right-hand figure shows the ordering for
$k_+<0$.  In both cases, the longitudinal momentum $k_+$ does not
    appear in the soft-quark line and, so, its intermediate-state
energy can never vanish. Consequently,
the energy denominators cannot vanish unless $\omega=\ell_+\ell_-$ is
    greater than or equal to $m_b^2$.

Finally, a light-front analysis of the denominators of diagram of
Fig.~\ref{fig:analyticity2}(d) yields
\begin{eqnarray}
&&2\pi  i \theta(-k_+) \frac{1}{\ell_+} \frac{1}{k_+}
\frac{1}{-k_++i\ve}
 \frac{1}{\bs{k}_\perp^2/k_+}
\frac{1}{\ell_- - (\bs{\ell}_\perp^2+m_b^2)/\ell_++i\ve}.
\end{eqnarray}
Again, the longitudinal momentum $k_+$ does not appear in the soft-quark
line, and, so, its intermediate-state energy can never
vanish. Consequently, this expression has a vanishing light-front energy
denominator only if $\omega=\ell_+\ell_-$ is greater than or equal
    to $m_b^2$.

From the foregoing examples in light-front perturbation theory, we
have seen that, owing to the unorthodox momentum routing in the soft
    function or to the presence of an $n_2$ Wilson line, a soft-quark
    line can carry an infinite $+$ longitudinal momentum without causing
    any particle line to move backward. This leads to a vanishing of the
    intermediate-state energy of the soft-quark line, and it is the
    mechanism by which a vanishing light-front energy denominator can
    appear for $\omega\geq 0$. Hence, in consequence of the $i\ve$ in
    the energy denominator, the soft function has a cut that runs from
    $\omega=-i\ve$ just below the real axis to infinity.  On the other
    hand, owing to the fact that none of the intermediate-state particle
    energies can be negative, energy denominators can never vanish if
    $\omega<0$, and the soft function is analytic everywhere in the
    complex $\omega$ plane except for the cut along the positive real axis.

%\nocite{*}

%\bibliography{covariant}% Produces the bibliography via BibTeX.

\end{document}